\documentclass[aps,twocolumn,showpacs]{revtex4}

\usepackage{amssymb,amsmath,graphicx,subfigure}

\begin{document}
\title{Heavy-tailed  targets  and (ab)normal  asymptotics in diffusive motion}
\author{Piotr Garbaczewski,  Vladimir Stephanovich and Dariusz K\c{e}dzierski }
\affiliation{Institute of Physics, University of Opole, 45-052 Opole, Poland}

\begin{abstract}
We investigate  temporal behavior   of probability density functions (pdfs) of    paradigmatic jump-type and continuous  processes  that, under
confining regimes,  share    common  heavy-tailed    asymptotic (target) pdfs. Namely, we have shown that under suitable confinement conditions,
the ordinary Fokker-Planck equation may  generate non-Gaussian heavy-tailed pdfs (like e.g.  Cauchy or  more  general  L\'{e}vy stable distribution)
in its long time asymptotics. For diffusion-type processes, our main focus is on their   transient  regimes  and specifically   the  crossover features,
when initially infinite number of the pdf moments drops down to a few or none at  all. The time-dependence  of the variance  (if in existence),  $\sim
t^{\gamma }$ with  $0<\gamma <2$, in principle  may be interpreted  as a signature  of sub-, normal or super-diffusive behavior  under confining
conditions; the exponent $\gamma $ is generically well defined  in  substantial  periods of time. However, there is no indication of  any universal time
rate hierarchy, due to a proper choice of the driver and/or external potential.
  \end{abstract}
\maketitle
\section{Introduction}

Among a large  variety of random walk and primordial noise models/choices, a distinguished  role is played by  heavy-tailed  symmetric
 L{\'e}vy-stable distributions,  commonly thought to be  an  exclusive  support  for so-called   L\'{e}vy flights. That view is founded on the concept of  independent and  identically-distributed random variables of  the jump-type and a broad class of infinitely divisible distributions,  that  are admitted by generalizing the familiar central limit theorem,  as encoded in the L\'{e}vy-Khintchine formula.  Stable densities generically have no second moments.

The  response of L\'{e}vy noise to  external potentials  is   most often  quantified by  means  of the  Langevin equation with  an additive  L\'{e}vy (stable)
  driver.  The  affiliated distribution functions (pdfs)  obey  so-called  fractional Fokker-Planck equations  where,   in order to handle  heavy-tailed  pdfs, the (Brownian case)  second spatial derivative is replaced by the  fractional one of order $0<\mu <2$,  c.f.  \cite{klafter}-\cite{gs1} and references therein. Under confining conditions, the  second  moments    of resultant  pdfs may exist  and then the temporal behavior of the variance may be employed to quantify  (on suitable time scales) sub-, or  super-diffusive features of the underlying   jump-type dynamics.

 In our  previous papers \cite{gs,gar,gs1} we have noticed that in addition to standard Langevin equation based methods,  an alternative  modeling approach is worth investigation. It is based on
 the concept of the L\'{e}vy-Schr\"{o}dinger semigroup-driven dynamics, \cite{olk,gs}.  Contrary to the  familiar
 (in the  context of  the  Brownian  motion)  mapping of the Fokker-Planck equation into the  Hamiltonian dynamical system,
 the non-Gaussian case makes a distinction between   these two    dynamical patterns of behavior. They  are inequivalent.

Our departure point was,  and still remains, the  "stochastic  targeting" (also named "reverse engineering")  strategy, \cite{klafter}:
given an invariant pdf $\rho _*(x)$, design a stochastic jump-type process for which that preselected density   is  a unique   {\it  asymptotic}  target.  In case of Langevin-driven processes, the basic reconstruction goal  amounts  to deducing  the   drift function of the process.

In Refs. \cite{gs,gar}, given the very same $\rho _*(x)$, we have
addressed  the existence issue  of a semigroup-driven dynamics  (e.g.
the fractional version of the generalized diffusion equation), which
 relies on the existence of a semigroup
potential ${\cal{V}}(x)=-\lambda \, (|\Delta|^{\mu/2}\rho _*^{1/2})(x)\, /\rho _*^{1/2}(x)$.  Since $\rho _*(x)$  is presumed to be shared with the  Langevin-driven L\'{e}vy process,
we know that it has a  non-Gibbsian   functional form, \cite{klafter}. Therefore standard thermalization and (ultimate)  thermal   equilibrium concepts
are   invalid in the L\'{e}vy   context.

In Ref.  \cite{gs1}  we  have  relaxed  the common pdf  constraint  and addressed  a
fully fledged reconstruction  problem for the semigroup dynamics:
given an invariant pdf, identify   the semigroup-driven  L\'{e}vy
process for which the  prescribed  pdf $\rho _*$  may
stand for a unique  asymptotic one.
 Since, to this end,   there is   no need to invoke the Langevin connection,  the  ensuing non-Gibbsian obstacle  does not appear  anymore.

Indeed, under new premises,    asymptotic pdfs  in the Gibbs form  are admissible and a class of
jump-type processes, non-trivially   responding   to environmental inhomogeneities,
becomes largely extended to pdfs that  are definitely related to   Gibbsian thermal equilibria,  \cite{gs1}.

The last observation suggests a possibility of  a major recasting of the original "reverse engineering"  problem
of Ref.  \cite{klafter} which was designed to handle jump-type processes only. Namely, while before we have set common
invariant pdfs for various jump-type processes,  presently  we shall  consider invariant  pdfs  that  are shared
 by, seemingly disparate,   jump-type (discontinuous) and   diffusion-type (continuous)   processes.
Invariant pdfs that show up  a power-law behavior (being e.g. of the inverse polynomial form)  are here allowed  and thus an issue
 of a  proper thermalization framework (Gibbsian equilibria), that encompasses heavy-tailed distributions,  reappears again.

An  issue disregarded in the past  \cite{montroll} is   that generic L\'{e}vy pdfs (like e.g. the familiar Cauchy distribution),
 plainly  against  casual views, may be embedded in suitable "exponential families" of pdfs  \cite{kapur,naudts,gs2}.  A careful exploitation
    of   standard   (Shannon) entropy extremum principles, \cite{kapur,gs2},
      allows to single out a  concrete  L\'{e}vy pdf as a  specific    exponential family  member, at a uniquely defined  inverse temperature  value.

   Anticipating further discussion, let us point out that our major observations  are quite general and  refer  to a broad class of pdfs that
    are   associated with symmetric stable noises and   their perturbed (confining regime) versions.
     It is only an  analytical and numerical tractability  reason, that makes  us mostly to  refer to the Cauchy driver  in the present paper.

    We note that in the  heavy-tailed  asymptotic regime  of  diffusion-type processes, the process remains continuous   and
    there is not (albeit appealing) "switch" to any   jumping  scenario.
     For all  times we deal with a  diffusion proceeding  in conservative force fields, whose pdf asymptotic  features  ultimately  appear
      to mimic those normally attributed to jump-type processes.

      In  a slightly careless manner one may think of a transition "from a diffusive motion  to  the L\'{e}vy flight behavior".  The corresponding  transient phenomena (that have been literally interpreted as a transition from a diffusive to jump-type  motion) were  experimentally  recorded in the past \cite{bronstein}. The diffusion vs jump scenario  interpretation  issue has appeared as well in the semiclassical description of so-called optical lattices, see e.g.  \cite{salomon}-\cite{renzoni}.

Apart from an  obvious  possibility to  regard  the L\'{e}vy  driver  (and thus  to invoke the  corresponding  non-Gaussian
   probability distribution)  as the major  random displacement  mechanism,  it is perhaps less obvious  that the Wiener driver
   actually may   give rise to  heavy-tailed pdfs  as well.   Then we  encounter   the dynamically generated  intermediate and/or transitional regimes, where the number of  the pdf moments asymptotically drops
down to a finite number or none at all.

   To elucidate these points,  here we   employ   as
    a toy model (albeit directly related to the previously mentioned optical lattice issue) a  one-parameter  family of Cauchy power
      pdfs, \cite{kapur}:
\begin{equation}\label{cf}
\rho_\alpha (x)=\frac{\Gamma(\alpha)}{\sqrt{\pi}\Gamma(\alpha -
1/2)}\frac{1}{(1+x^2)^\alpha},\ \alpha>1/2
\end{equation}
as  a   reference exponential family of pdfs   (see e.g. \cite{kapur})  that   comprises the classic  Cauchy distribution as its   member.
This familiy is an exponential one due to a trivial transformatin into the Gibbs-loking function with a logarthmic exponent. (Actually, for each "canonical" L\'{e}vy pdf an analogous exponential family embedding can be accomplished.)

The above Cauchy   family  arises naturally via the  \it standard \rm  maximum entropy  principle.
Namely, one seeks an extremum of the (dimensionless) Shannon entropy
$S(\rho)=-\int\rho \ln \rho dx$  of a continuous probability distribution $\rho(x)$  under the constraint that
   the expectation value $<\ln(1+x^2)>$  takes
an a priori  prescribed  value, \cite{kapur}.

 If we regard a  logarithmic   function $\alpha \ln(1+x^2)$ as  an external force potential, with  $- 2 \alpha  x/(1+x^2)$  being   interpreted as the  forward drift in
  the Langevin equation, then we end up with a familiar   Fokker-Planck equation   (Wiener driver in action).

   A functional form  of  the force term is identical to that of the  physically motivated "cooling force" in optical lattice discussions \cite{lutz}.
With that  restoring force, we are capable of  generating all   members of the   above  Cauchy family  as target pdfs   in the large
  time asymptotic of  diffusion-type  processes.
   This linear Fokker-Planck equation option has  appeared in the literature before, \cite{lutz,lutz1},
    also in the attempt to interpret the Cauchy family as a family of Tsallis distribution functions \cite{stariolo}.

 All members of the   Cauchy  family  can be  achieved as   asymptotic targets for    jump-type  processes. Both  the L\'{e}vy-Langevin and L\'{e}vy
  semigroup driven dynamics  may be employed to this end.  In turn, the same "targeting" is valid if we resort to diffusion-type processes.

\begin{figure}
\begin{center}
\includegraphics [width=0.9\columnwidth]{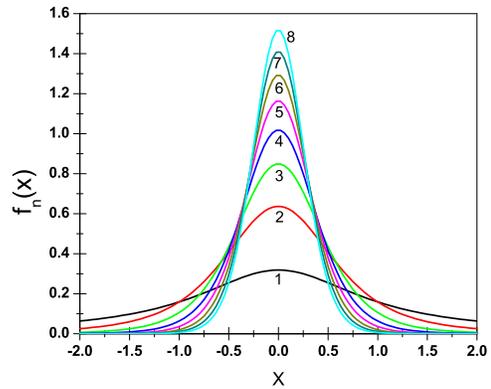}
\end{center}
\caption{ The members of Cauchy pdf family for $1\leq  n \leq 8$. } \label{fig:jed}
\end{figure}

An important  intrinsic    property  of the Cauchy family  has to do with a number of moments of those pdfs. For a while, let us consider a  parameter $\alpha$ in Eq. \eqref{cf} to assume  integer values only:  $\alpha \to n\geq 1 $. Then one  immediately observes that
the  Cauchy hierarchy  of pdfs, c.f. Fig.~\ref{fig:jed}, involves a monotonically  growing number (as parameter $n$ grows from $n=1$  to infinity)
of  moments of the probability distribution, beginning from none at all for $n=1$. The growing number of moments in existence amounts to an
improvement  of the pdf localization and entails  increasing strength of  confinement of Cauchy jumps (e.g.  flights), c.f. also \cite{gs,gar}, if the Cauchy driver is  assumed to be in action.

   An  appealing feature of the Cauchy family  is that, even if looking exceptional   in many respects, it has  received an ample literature  coverage  not only  in the semiclassical analysis of optical lattices \cite{zoller}-\cite{renzoni}.
In below,  we shall elaborate a bit on another physically important  context  of Tsallis entropies and related (via entropy  maximum principles) Tsallis pdfs.

Indeed, Cauchy family pdfs    were discussed in search for  a  statistical  and thermal (equilibrium)   foundation   for Tsallis entropies and   associated with them probability distributions. That includes asymptotic properties of various Fokker-Planck equations,  nonlinear   being included, as a dynamical justification for them, see e.g.   Refs. \cite{tsallis}-\cite{barkai}.
 From a thermal equilibrium  point of view, in reference to heavy-tailed pdfs,  our approach appears to provide an alternative proposal to that originally  formulated  in  Refs. \cite{alemany,alemany1,tsallis}  and \cite{stariolo,borland}.

Subsequently we shall discuss in some detail various random motion scenarios that may generate the above Cauchy family of pdfs in their
large time asymptotics. After identifying the  involved Cauchy semigroup dynamics,   we shall demonstrate that  diffusion-type processes  as well generate heavy-tailed  (Cauchy, to be specific) pdfs, in affinity with the case of jump-type processes described by fractional  Fokker-Planck equations.

Somewhat surprisingly, it turns out that  the conventional diffusive dynamics  may  be  regarded as an alternative (to  the fractional one)
dynamical  model of random behavior that    produces non-Gaussian pdfs. This includes heavy tailed ones,  that are casually thought to
arise exclusively in relation to    "free" (we stress  the role of independent, identically distributed random displacements) jump-type random motions.

 In   fact, we demonstrate  the  validity of the diffusive   relaxation scenario  from an initial $\delta $ or a Gaussian function
 to virtually {{\em any}} member of  Cauchy power pdf hierarchy \eqref{cf}.  In the diffusive dynamics, the resultant
pdfs appear to have the Gibbs - Boltzmann form  (as  should be for the exponential family of pdfs).
All those observations  readily  extend to a   broad variety of  non-Gaussian pdfs that have been   invented  in connection with
 the concept  of   confined  L\'{e}vy flights.

Our discussion  uncovers an interplay between L\'{e}vy and diffusion-type processes that has been left
 basically ignored  in  the current "stochastic modeling" literature. We point out that, in principle,   one may even not be able
 to discriminate between  these two classes   (continuous vs discontinuous) of random processes, if  in the vicinity
 of respective (common for both) equilibrium states.
Such observation may be particularly useful to avoid erroneous interpretation, in cases when  an experimentally founded data analysis
indicates a transient dynamics with   a (possibly unusual) crossover behavior, like e.g. that  reported   in Ref. \cite{bronstein}.

In the present paper  Langevin noises (L\'{e}vy and Wiener driver alike) are considered as additive. For a completeness of exposition, we
  point out that there are reports in the literature that Langevin equations with   the multiplicative (Wiener) noise do  give rise to heavy-tailed distributions, see e.g. Ref. \cite{manor}.

\section {Entropy extremum principles}

\subsection{Shannon vs Tsallis entropies}

An influential  Ref. \cite{montroll} appears to have harmed  an open-minded approach to extremum entropy principles, with regard to heavy-tailed probability distribution functions. Its  main  message has been often  uncritically repeated: "the derivation of distributions  with inverse power tails from   a maximum entropy formalism   would be a consequence  only of an unconventional auxiliary condition that involves   the specification of the average value of a  complicated function". This statement has been amplified at the end of Section 2 of Ref. \cite{montroll}:
"It is difficult to imagine that  anyone in an a priori manner  would introduce a set of auxiliary conditions  that   could yield    the logarithmic term  that   appears in the (Shannon) entropy function associated   with the L\'{e}vy distribution."
In Ref. \cite{tsallis}, the pertinent formula (their Eq. (7)) is called the  "ad hoc constraint".

A possible way out of this apparent difficulty has been proposed in the past. It amounts to  abandoning  the standard Shannon entropy usage  in extremum principles  and deviate towards non-extensive   generalizations  of the notion of entropy (effectively one ends up with a non-extensive thermostatistics), so that  the heavy-tailed distribution would "more or less naturally" arise.  That has been suggested in Refs. \cite{tsallis,alemany,alemany1} and  followed in the interpretation of experimental data in Refs. \cite{bronstein}-\cite{renzoni}.

  We wish to demonstrate that  the non-extensive proposal is  \it not \rm  the only admissible route, if one looks for mechanisms that generate  heavy-tailed target pdfs.  The conventional Shannon entropy strategy appears to work  satisfactorily as well.

If one  seeks  for  an extremum of the Shannon entropy
$S(\rho)=-k_B\int\rho \ln  ( \sigma \rho )dx$  of a continuous probability distribution  ($k_B$ is Boltzmann constant and parameter $\sigma >0$ determines the characteristic length of a system),  under the constraint that
 the expectation value $\int x^2 \rho (x) dx= \sigma ^2$ \cite{tsallis}, an automatic outcome is a Gauss density   $\rho _*(x)= (\beta /\pi )^{1/2}\,  \exp(-\beta  x^2)$,  where  an  (implicit) Lagrange multiplier $\beta $      takes the  value $\beta = 1/2\sigma ^2$. Finally, the presupposed   Gibbs thermal  equilibrium condition  imposes $\beta  =1/k_BT$, \cite{tsallis,landsberg}.

   Let us indicate that for generic Gibbsian  densities  $\rho _*(x)=(1/Z)  \exp(-\beta V(x)$, where $V(x)$ stands for an external potential and $Z$  is a normalization factor,  the  previous  moment constraint corresponding to the Lagrange multiplier $\beta $  may  readily be  generalized to the form
    $\langle V \rangle =const $  whose  equivalent form  is $\langle \ln \rho _* \rangle = const'$.

So called "ad hoc" constraints of Refs. \cite{montroll} and \cite{tsallis}, albeit with no reference to any effective potential, nor exponential  (Gibbs-looking) recasting  of the  involved pdfs, are nothing more than the demand
 \begin{equation}
\langle \ln \rho _* \rangle =   \int \rho (x) \ln \rho _*(x) dx  = const
\end{equation}
where $\rho _* $ stands for  \it  any \rm  a priori pre-selected  L\'{e}vy stable pdf.  C.f. Eq. (21) in Ref. \cite{montroll} and Eq. (7) in Ref. \cite{tsallis}.  We emphasize that the pertinent L\'{e}vy distributions are "free noise"   models and have nothing to do with the notion of confined L\'{e}vy flights.

The proposal of Refs. \cite{alemany,alemany1,tsallis} amounts to considering the new entropy   function
\begin{equation}
 S_q[\rho ] = {\frac{k_B}{q-1}}  \left(1-\int d(x/\sigma ) [\sigma\rho (x)]^q\right )
 \end{equation}
with a real parameter $q$.  An optimization  \cite{caceres} of a suitable likelihood function, under a constraint \begin{equation}
\langle x^2\rangle _q =  \int d(x/\sigma ) x^2 \, [\sigma \rho (x)]^q = \sigma ^2
\end{equation}
results in an extremal pdf:
\begin{equation}
\rho _q(x)= {\frac{1}{Z_q}}\, \left[ 1-  \beta (1-q) x^2\right] ^{1/1-q}
\end{equation}
where
\begin{equation}
Z_q = \left[ {\frac{\beta (q-1)}{\pi }}\right]^{1/2}  {\frac{\Gamma (1/(q-1))}{\Gamma ((3-q)/2(q-1))}}
\end{equation}
and we need $1< q< 3$ to secure convergence of the normalization integral.

We note that, in view of a direct $q\rightarrow 1$  connection   with the  Shannon entropy (and the  Gibbs density), an interpretation $\beta =1/k_BT$ is enforced.  That, in turn, has been a starting point in Refs. \cite{stariolo,borland} to deduce the Tsallis pdf in an asymptotic regime of a well defined Fokker-Planck dynamics.  The resultant   invariant pdf has   the Gibbs form $\rho _q(x)= \exp [-\beta V(x)]/Z_q$,
   provided  the "external force" potential
reads:
\begin{equation}
V(x)=  {\frac{1}{\beta  (q-1)}}  \ln  [1+\beta (q-1)x^2]\, .
\end{equation}

Let us redefine the involved constants. Namely, after \cite{lutz}  (we  modify  Eqs. (4) therein),  let us set
\begin{equation}
\beta = {\frac {\alpha }{2D}}, \quad q=  1 + {\frac{2D}{\sigma ^2\alpha }}.
\end{equation}

That results in the $V(x)= \sigma ^2 [ 1+(x/\sigma)^2]$  and $\rho _q(x)$ effectively turn over into:
\begin{equation}
\rho _q(x)= {\frac{1}{Z_q}} \,  [1+ (x/\sigma )^2]^{- \beta \sigma ^2},
\end{equation}
which differs from Eq. \eqref{cf} by
 a  trivial replacement of the exponent $\alpha $ by $\beta \sigma ^2$  and an explicit usage of dimensionless argument $x/\sigma $.  Clearly, if $\beta $ plays the role of  $1/k_BT$, the    signature of thermal equilibrium needs to be directly transferrable to the exponent $\alpha $ as well.

\subsection{Shannon entropy extremum for random systems in logarithmic potentials}

Cauchy family pdfs are labeled by a continuous parameter $\alpha >1/2$, whose physical meaning seems to be obscure.   In addition to considerations of the previous subsection we shall give more  arguments to the contrary. Even without explicitly  relying on the optical lattice contexts, \cite{zoller}-\cite{renzoni}.

At this point we invoke  a  classification of maximum entropy principles (MEP) as given in Ref.~\cite{kapur}. Let us
look for pdfs that derive from so-called first inverse MEP:  given a  pdf $\rho (x)$, choose
an \it appropriate \rm set of constraints such that  $\rho (x)$ is  obtained if Shannon  measure of  entropy  is maximized (strictly speaking, extremized) subject to those constraints.

Namely if a system evolves in a potential
$V(x)$ (at the moment,  we consider a coordinate $x$ to be dimensionless), we can introduce the following functional
\begin{equation} \label{fx1}
L \{ \rho (x) \}=-\lambda \int_{-\infty}^\infty V(x) \rho (x) dx -  \int_{-\infty}^\infty \rho (x) \ln [\rho (x)]dx\, .
\end{equation}
 The first term comprises  the mean value of a potential.
 A constant $\lambda$ is (as yet  physically unidentified) Lagrange multiplier, which takes care of aforementioned constraints. The  second term   stands for  Shannon entropy of a continuous (dimensionless) pdf $\rho (x)$.
  An extremum of the functional $L \{ \rho (x) \}$  can be found by means of standard variational arguments and gives rise to  the
   following  general form of an   extremizing  pdf $\rho _*(x)$: $ \label{fx2} \rho _*(x)=C\exp(-\lambda
   V(x))$
which, if regarded as the Gibbs-Boltzmann pdf,   implies that the  parameter $\lambda$ can be interpreted  as inverse temperature,
$\lambda=(k_BT)^{-1}$,  at which a state of equilibrium (asymptotic pdf) is reached by a
  random dynamical  system in a confining potential   $V(x)$.

A deceivingly simple question has been posed in chap. 8.2.4 of Ref. \cite{kapur}.   Having dimensionless logarithmic
potential  ${\cal{V}}(x)= \ln (1+x^2)$, one should  begin with evaluating a mean value
 ${\cal{U}}= \langle {\cal{V}}\rangle$  $\equiv \int_{-\infty}^\infty {\cal{V}}(x) \rho (x) dx$.  Next one needs to  show that
only if \it  this particular value \rm  is prescribed in the above MEP procedure,  Cauchy distribution will ultimately arise.
Additionally,  one should  answer  what kind of distribution would arise if \it  any other  \rm  positive expectation value is chosen.
The answer proves not  to be that  straightforward and we shall analyze this issue below.

To handle the problem  we admit all  pdfs $\rho (x)$ for which the mean value $\langle \ln (1+x^2) \rangle $ exists,
 i.e. takes \it  whatever  \rm  finite positive value.  Then,  we adopt  the previous  variational procedure with the use of
Lagrange multipliers.  This procedure shows that what we extremize is not the (Shannon) entropy itself,
but a functional  ${\cal{F}}$ with a clear thermodynamic connotation  (Helmholtz free energy analog):
 \begin{equation} \label{fe1}
\Phi  (x) = \alpha \,   {\cal{V}}(x) + \ln \rho (x)  \rightarrow
{\cal{F}} = \langle \Phi \rangle  = \alpha \, \langle
{\cal{V}}\rangle - {\cal{S}}(\rho )\, .
\end{equation}
Here ${\cal{S}}(\rho )= - \langle \ln \rho \rangle $ and $\alpha $
is a Lagrange multiplier. From now on,  we consider  an exponent
$\alpha$ in the form
\begin{equation}
 \alpha=\epsilon_0/(k_BT)
\end{equation}
 where $\epsilon_0$
is a  characteristic energy scale of a system.  Note that here we
encounter  a dimensionless
 version of   a familiar formula  $F=U-TS$,   relating  the Helmholtz free energy $F$,
internal energy $U$ and entropy $S$ of a random dynamical system.

The extremum condition $\delta {\cal{F}}(\rho )/ \delta \rho =0$
yields an extremizing pdf in the form
  $\rho _{\alpha  }(x) = (1/Z_{\alpha })\, (1+x^2)^{-\alpha }$,
provided the normalization factor  $Z_{\alpha } = \int_{-\infty}^\infty (1+x^2)^{-\alpha
}\, dx $ exists.  It turns out that the integral can be evaluated explicitly in terms of
$\Gamma$ - functions  for all $\alpha >1/2$,   ending up at the previously introduced one-parameter Cauchy family of pdfs \eqref{cf}.

To complete an extremum procedure we can in principle deduce  a
numerical  value of the (Lagrange)  parameter $\alpha $,  by resorting
to our  assumption that the mean value $\langle {\cal{V}} \rangle
_{\alpha }$ has  actually  been a priori  fixed at a concrete value. This route is not at all straightforward.

To identify  the values of above  $\alpha$, we need an explicit expression for the  mean value
\begin{equation}\label{ln1}
 {\cal{U}}_{\alpha }=  \langle {\cal{V}}\rangle _{\alpha }  =
\frac{\Gamma(\alpha)}{\sqrt{\pi}\Gamma(\alpha - 1/2)}\int_{-\infty}^{\infty}\frac{\ln(1+x^2)}{(1+x^2)^\alpha}dx.
\end{equation}
It turns out that it can  be given  in terms of the digamma function $\psi(x)=d(\ln \Gamma)/dx$:
\begin{equation}
{\cal{U}}_{\alpha}=
-\frac{2\pi}{\sin(2\pi\alpha)}+\psi(1-\alpha)-\psi\left(\frac 32-\alpha\right),\ \alpha>\frac 12.\label{ln3}
\end{equation}
This function is divergent at
$\alpha =1/2$ (see also below) and decays monotonously at large $\alpha$.
This decay  is conrolled by an asymptotic expansion
\begin{equation}\label{as}
{\cal{U}}_{\alpha}  \approx   \frac{1}{2\alpha}+\frac{3}{8\alpha^2}+\frac{1}{4\alpha^3}+...
\end{equation}
  The decay of ${\cal{U}}_{\alpha}$ at large $\alpha$  obeys the inverse power law. It follows that the expansion \eqref{as} actually  gives
a very good approximation of ${\cal{U}}_{\alpha}$ for $\alpha >3$.

We note that, apparently,  Eq. \eqref{ln3}  involves another  divergence problem,  if we choose   integer  $\alpha $.
This obstacle can be circumvented by transforming Eq. \eqref{ln3} to an equivalent form that has no
 (effectively  removable) divergencies. Namely,  we get
\begin{equation} \label{gl3}
{\cal{U}}_{\alpha}=-\pi \tan \pi \alpha +\psi(\alpha)-\psi\left(\frac 32-\alpha\right)
\end{equation}
and the  tangent  contribution vanishes for integer $\alpha $.
 On the other hand,    this expression shows that the divergence of ${\cal{U}}_{\alpha}$ at $\alpha \to 1/2$  originates
  from the first term in \eqref{gl3},  as $\psi$ functions have finite values at
this point. Near $\alpha =1/2$ the first term of Eq. \eqref{gl3} diverges as $(\alpha-1/2)^{-1}$.

\begin{figure}
\begin{center}
\includegraphics [width=0.9\columnwidth]{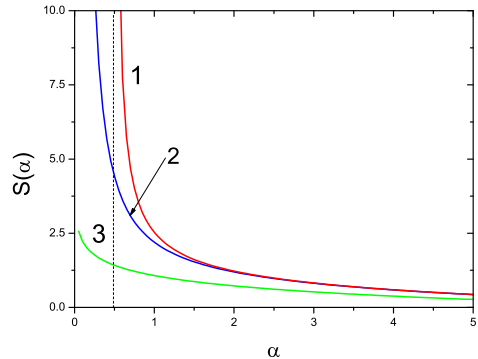}
\end{center}
\caption{ ${\cal{S}}_{\alpha }$  for Cauchy family (curve 1) and its asymptotic expansion at large $\alpha$  (curve 2) .
${\cal{S}}^G_{\alpha }$ for Gaussian family  is also shown (curve 3).}\label{fig:sei}
\end{figure}

With an explicit expression for   Cauchy family pdfs in hands,  we readily  evaluate  Shannon entropy to obtain
\begin{equation}
{\cal{S}}_{\alpha }=-\int_{-\infty}^{\infty}\rho_\alpha (x)\ln \rho_\alpha (x)dx=
  \ln Z_\alpha+\alpha \,  {\cal{U}}_{\alpha}. \label{sa}
\end{equation}
Then, the  (as yet dimensionless) Helmholtz free energy ${\cal F}_{\alpha }$   reads
\begin{equation} \label{fedim}
{\cal {F}}_{\alpha }=   \alpha {\cal{U}}_{\alpha }  -  {\cal{S}}_{\alpha } \equiv -\ln Z_{\alpha } ,
\end{equation}
with $\alpha $  being the dimensionless analog of the  inverse temperature.

We note, that in view of the divergence of $Z_{\alpha }$, both the Shannon entropy and the Helmholtz  free energy
 (likewise ${\cal{U}}_{\alpha }$) cease to exist
at $\alpha = 1/2$.
We plot  ${\cal{S}}_{\alpha }$   as a function of $\alpha $ in Fig.\ref{fig:sei}. It is seen that entropy  monotonously
 decays for $\alpha >1/2$ and for larger values
of $\alpha $. An   asymptotic expansion of the entropy   shows  logarithmic plus inverse power signatures
\begin{equation}\label{sas}
{\cal{S}}_{\alpha } \approx \frac 12\left(1-\ln \frac{\alpha}{\pi}\right) +\frac{3}{4\alpha}+\frac{3}{8\alpha^2}+...
\end{equation}
These series are shown along with the  entropy in Fig.~\ref{fig:sei}.

As $\alpha $   grows, the number of moments of respective pdfs increases. That allows to expect that  an "almost Gaussian" behavior should be displayed by $\alpha \gg 1$ members of Cauchy family. This is indeed the case as discussed in Section II.G.

\begin{figure*}
\begin{center}
\includegraphics [width=0.66\columnwidth]{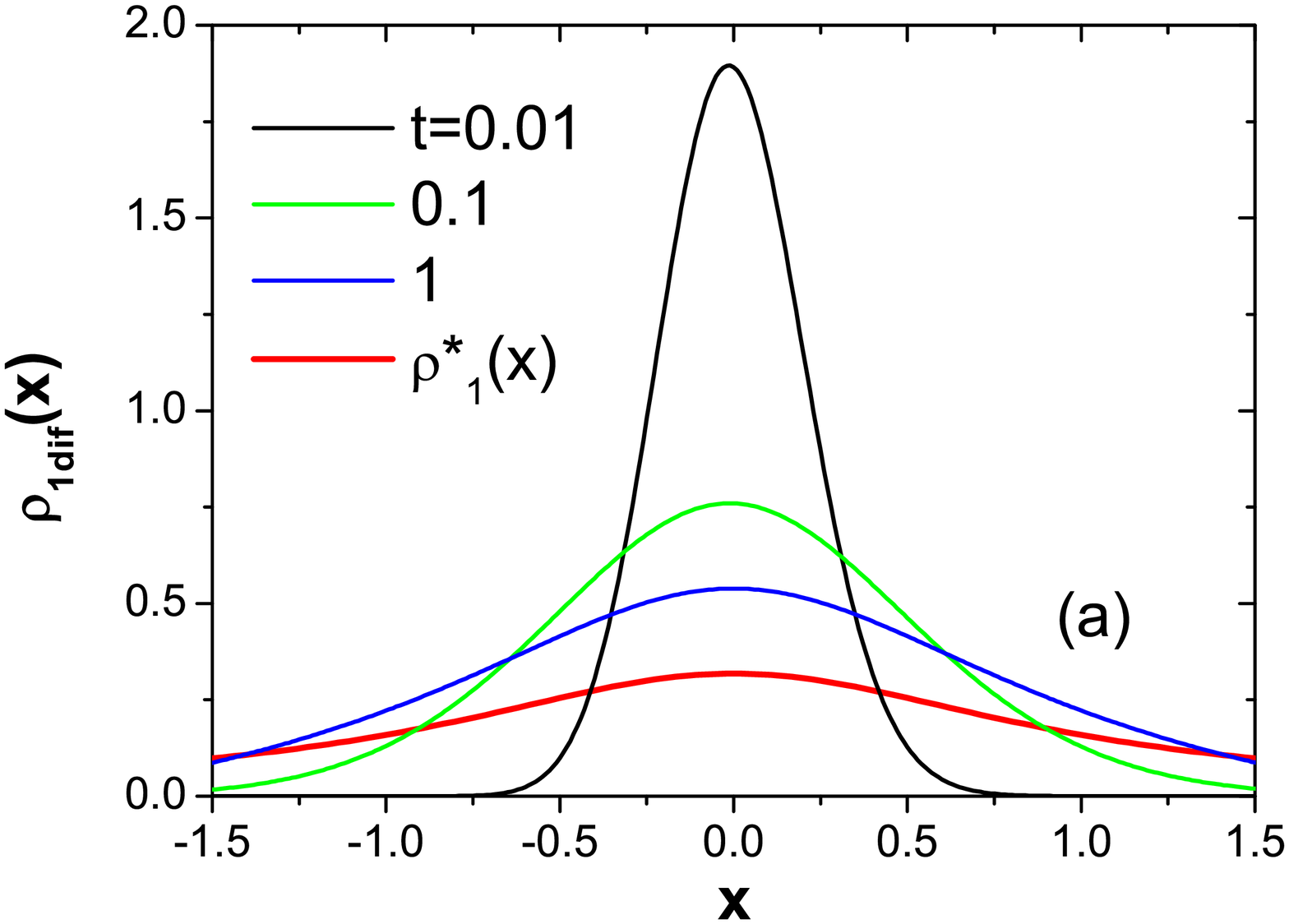}
\includegraphics [width=0.66\columnwidth]{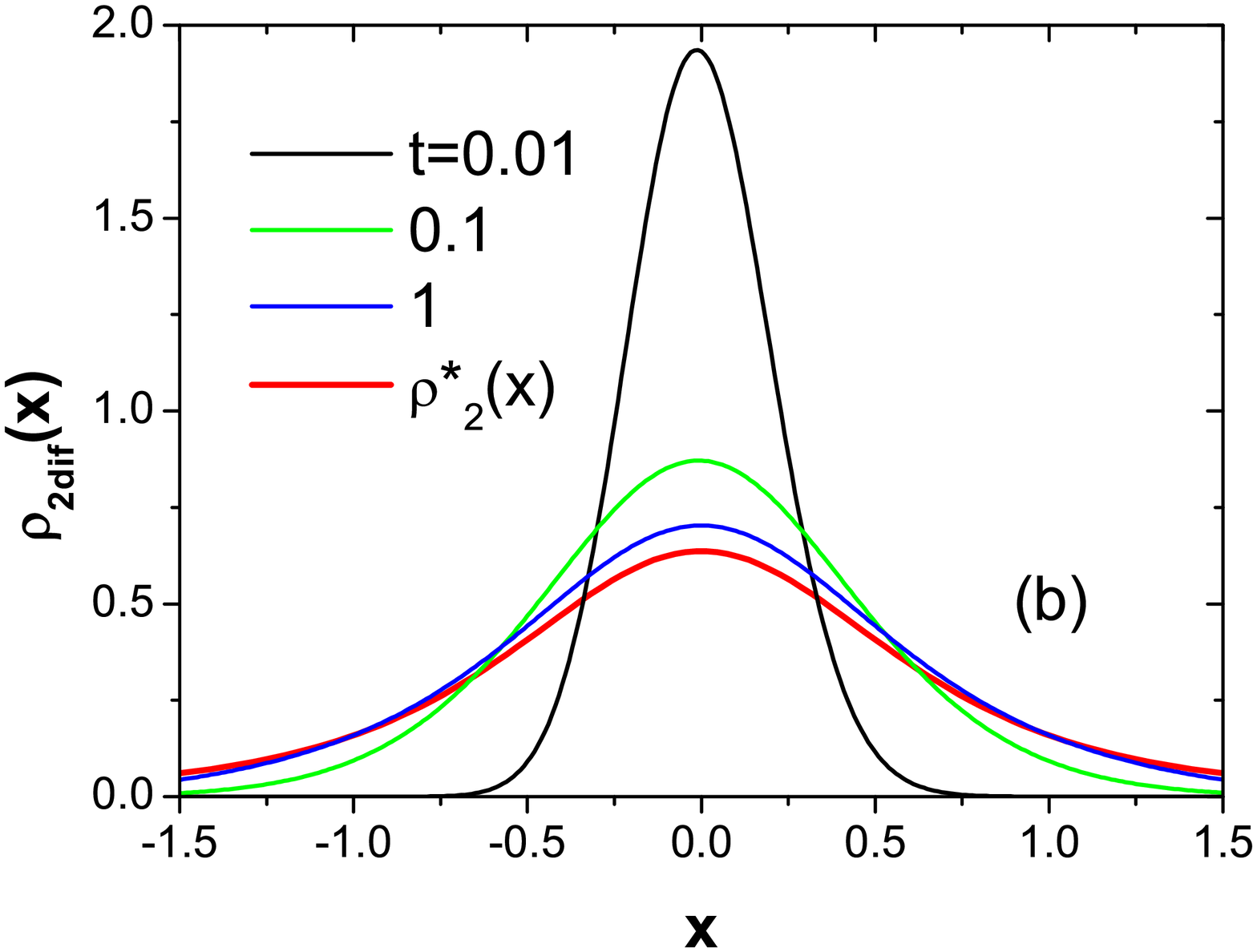}
\includegraphics [width=0.66\columnwidth]{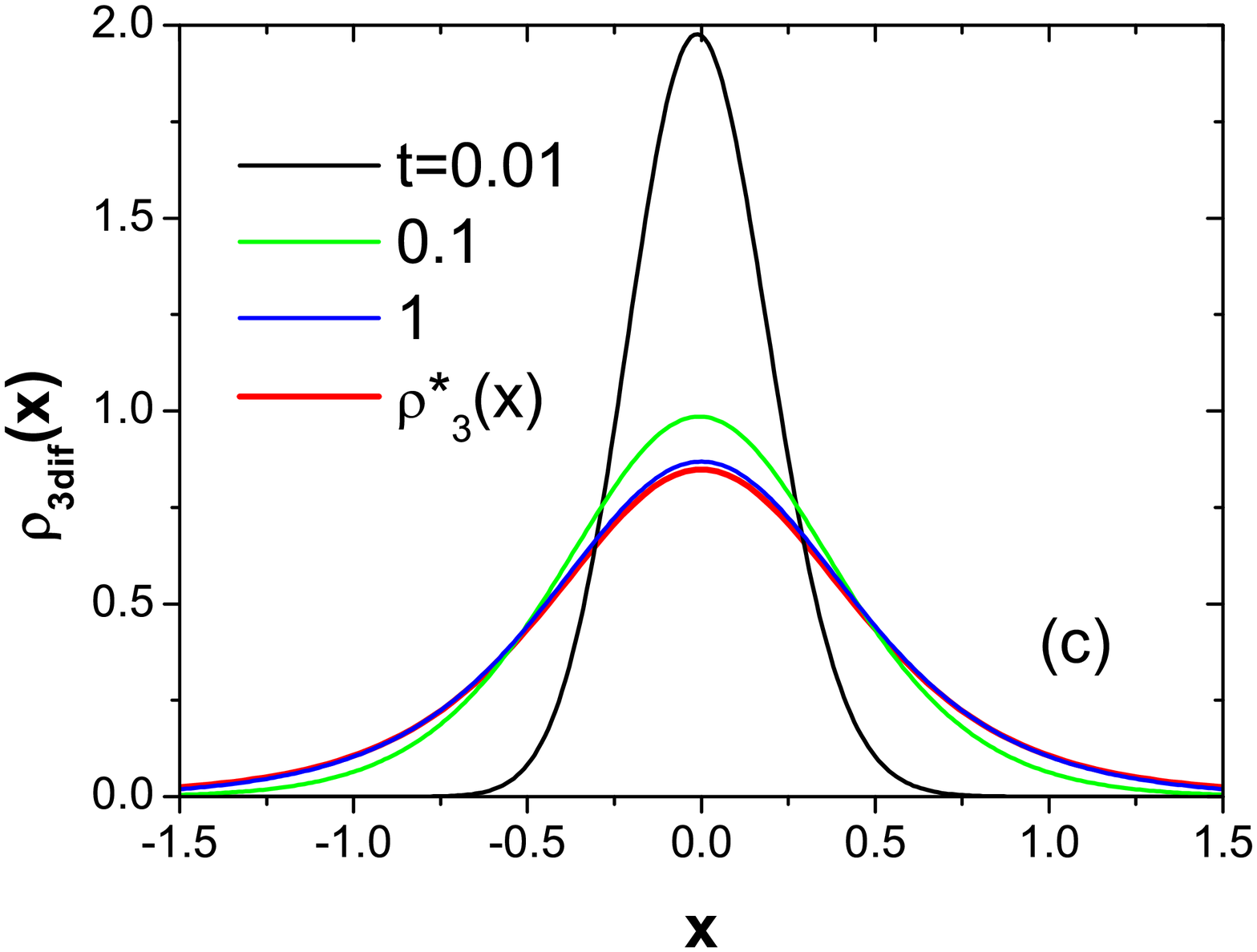}
\end{center}
\caption{Time evolution of pdf's $\rho(x,t)$ for Smoluchowski  processes in logarithmic potential
$\ln(1+x^2)$. The initial ($t=0$) pdf is set
to be  a Gaussian with height 25 and half-width $\sim 10^{-3}$. The first depicted stage of evolution corresponds to  $t=0.01$.  Target pdfs are the members of Cauchy family for $\alpha=1$ (panel (a)), 2 (panel (b))  and 3 (panel (c)) respectively.}\label{fig:zu}
\end{figure*}

\subsection{Thermalization via  Fokker-Planck dynamics}

Let us now show that the  variational principle  explicitly
identifies an  equilibrium solution of the Fokker-Planck equation for standard Smoluchowski diffusion processes.
 To address our thermalization issue correctly, we now use
dimensional units. The Fokker - Planck equation that drives an initial   probability density $\rho(x,t=0)$ to its final (equilibrium) form $\rho(x,t \to \infty)$  reads
\begin{equation} \label{gpe}
\partial _t\rho = D\Delta \rho -  \nabla \cdot ( b(x) \rho ).
\end{equation}
Here, the drift field $b(x)$ is time-independent and conservative, $b(x)=-\nabla V(x)/(m\gamma)$ ($V(x)$ is a potential, while $m$ is a mass and $\gamma$ is a reciprocal relaxation time of a system).
We keep in mind that $\rho$ and $b\rho $ vanish at spatial infinities or other integration interval borders.

If Einstein fluctuation-dissipation relation $D=k_BT/m\gamma $ holds, the equation \eqref{gpe} can be identically rewritten
 in the form $\partial _t\rho = \nabla [\rho \nabla \Psi ]/(m\gamma)$, where
\begin{equation} \label{tyr0}
\Psi = V + k_BT \ln \rho
\end{equation}
whose mean value is indeed the Helmholtz free  energy  of  random  motion
\begin{equation} \label{tyr1}
F \equiv \left< \Psi \right> = U - T S.
\end{equation}
Here the  (Gibbs)  entropy reads $S =  k_B {\cal{S}}$, while   an
internal energy is $ U = \left< V\right>$. In view of assumed
boundary restrictions at spatial infinities, we have $ \dot{F}  = -
(m\gamma ) \left<{v}^2\right>  \leq 0 $ , where $v=  -\nabla \Psi
/(m\gamma)$.   Hence, $F$ decreases as a function of time  towards
its  minimum $F_*$, or  remains constant.

Let us consider the stationary (large time asymptotic) regime
associated with an invariant density $\rho _{*}$ (c.f.  Ref. \cite{mackey} for an extended discussion of that issue).
Then,  $\partial _t\rho = 0$  and we have $\nabla \Psi [\rho_*] =C \rho_*$ ($C$ is arbitrary constant) which yields  $
\rho _{*} = (1/Z)\,  \exp[ - V/k_BT]$.
Therefore, at equilibrium:
\begin{equation} \label{tyr3}
\Psi _{*} = V + k_BT \ln \rho _{*}  \Longrightarrow \langle \Psi _{*} \rangle =
 - k_BT \ln Z  \equiv  F_{*},
 \end{equation}
to be compared  with  Eq.~\eqref{fedim}. Here, the  partition function equals $Z= \int \exp(-V/k_BT) dx$, provided that the integral is convergent.
 Since  $Z= \exp (-F_*/k_BT)$ we have  recast  $\rho _*(x)$  in the familiar Gibbs-Boltzmann form  $\rho _* = \exp[(F_* - V)/k_BT]$.

On physical grounds, $V(x)$  carries dimensions of energy. Therefore to establish a physically justifiable
 thermodynamic picture of Smoluchowski diffusion processes, relaxing to Cauchy family pdfs  in their large time asymptotics,
  we need to assume that  logarithmic potentials  ${\cal{V}}(x)=\ln (1+x^2)$  are dimensionally scaled to the form
   $V(x)= \epsilon _0 {\cal{V}}$ where $\epsilon _0$ is an arbitrary constant with physical dimensions of energy.

By employing  $ (1/\epsilon _0)\, V(x) = \ln (1+x^2)$, we can recast  previous variational arguments (MEP procedure) in
terms of dimensional thermodynamical functions. Namely, in view of $k_BT \Phi (x)= V(x)  +k_BT\, \ln \rho (x)$ we have an
 obvious transformation of Eqs.~\eqref{fe1} and \eqref{fedim} into Eqs.~\eqref{tyr1} and \eqref{tyr3}
  respectively with  $ k_BT \Phi (x)= \Psi (x)$.

 The above dimensional arguments tell us that in  confining logarithmic potentials, Cauchy family of pdfs  can be regarded as a one-parameter family of  \it equilibrium \rm  pdfs, where the reservoir temperature $T$ enters through the exponent $\alpha $. Proceeding in this vein,  we note that  $\epsilon _0$ should be regarded as a characteristic energy (energy scale) of  the considered  random system.

We observe that   $\alpha \to \infty$  corresponds  to $T \to 0$ i.e. a maximal localization (Dirac delta limit)
 of the corresponding pdf.
The opposite limiting case $\alpha \to 1/2$ looks interesting. Namely, we have
${\cal{S}}(\alpha \to 1/2)\to\infty$. To grasp  the meaning of this limiting regime, we rewrite $\alpha =1/2$ in the form
$k_BT=2\epsilon _0$.  Accordingly, the  temperature scale, within which our system may at all be set at  thermal equilibrium,
 is  bounded:  $0<k_BT<2 \epsilon_0$.
 For temperatures exceeding the  upper bound  $T_{max} = 2\epsilon _0/k_B$  {\em {no thermal equilibrium}} is possible in
  the presence of  (weakly, i.e. weaker then, e.g., $V(x)\sim x^2$ ) confining  logarithmic potentials
   $V(x)=  \epsilon _0\,  \ln (1+x^2)$. The case of $\alpha = 1$ i.e. $k_BT=\epsilon _0$ corresponds
to Cauchy density.

For clarity of presentation
  (dynamical interpolation scenarios between initial Gaussian and equilibrium Cauchy-type pdfs
do not seem to have ever been  explicitly  considered in the literature), in Fig. \ref{fig:zu} we plot various stages of the
diffusive Fokker - Planck dynamics for processes that  all have been  started from a narrow Gaussian.
  The resultant (large time asymptotic)  equilibrium  pdfs  are members of the Cauchy family,
   labeled  respectively by  $\alpha=1$, $2$ and $3$.

\section{Cauchy family targeting: transient dynamics and relaxation patterns}
To make the paper self-contained,  here  we describe  the dynamics of probability distributions  $\rho(x,t)$, whose   time evolution
 is started from the  Dirac delta-type initial data (actually, a very narrow Gaussian),  but  might be driven by  three \it different \rm
stochastic mechanisms.  Apart form the common  initial data choice, we shall  demonstrate that  a common target (invariant) pdf
 to which the pertinent processes relax in the large time asymptotic,  is  shared   by:\\
  (i) diffusion-type   process with a well defined drift function $b_{drift}(x)$ and the Wiener driver in action,\\
  (ii)  Langevin-driven  jump type process,  with a  suitable drift function  $b_{jump}(x)$  and the Cauchy driver in action,\\
  (iii)  L\'{e}vy-Schr\"{o}dinger  semigroup-driven  jump type process  of Ref.~\cite{gs},  with an (effective)  semigroup
   potential ${\cal{V}}(x)$ and  the Cauchy driver implicit.\\
 The above three patterns of dynamical behavior  will be demonstrated  by assuming that  asymptotic (target) pdfs   are members of the   Cauchy  family
\begin{equation}\label{ca1}
\rho_{*n}=\frac{A_n}{(1+x^2)^n},\ A_n=\frac{(n-1)!}{\sqrt{\pi}\Gamma(n-1/2)}.
\end{equation}
It turns out that for our purposes it suffices to investigate the above dynamics for $1\leq n\leq 4$ only, the dynamics for higher $n$
 being a repetition of  the established patterns of behavior.

 Since we require   processes  (i)-(iii)  to share an asymptotic pdf $\rho_{*n}$ for each choice of $n$, the  semigroup  potentials
 (and drift functions, if applicable) will certainly  differ from case to case. An additional  technical input that  needs to
 be mentioned in  view of a common initial pdf (actually, an "almost" Dirac delta) for all
  considered processes,  is that we refer to the well developed theory of  so-called Schr\"{o}dinger boundary data
  and stochastic interpolation  problem, \cite{zambrini,olk}.
  That allows to devise   Markovian dynamics scenarios which interpolate between a priori prescribed  initial and terminal (eventually  those referring to  the large time asymptotic and relaxation regimes) pdf data.

\subsection{Cauchy target, $\rho_{*1}$.}
Here and subsequently, for notational and computational simplicity,  we shall get rid of all freely adjustable   parameters
(like e.g.  $\beta$, $\gamma$, see e.g. \cite{gs})  and set them equal one. The  familiar Cauchy pdf:
\begin{equation}\label{ca2}
\rho_{*1} =\frac{1}{\pi(1+x^2)}.
\end{equation}
is known to be  a stationary pdf of the Ornstein-Uhlenbeck-Cauchy (OUC)  stochastic  process, \cite{olk,gs}.

\subsubsection{Cauchy driver: jump-type process}
Following a  widely accepted reasoning  we take  for granted that the
Langevin equation, with additive deterministic and  L\'{e}vy
noise terms, gives rise to the  fractional Fokker-Planck
equation, whose form  faithfully  parallels  the Brownian  version: $
\dot{x}= b(x)  + A^{\mu }(t)
\Longrightarrow \partial _t\rho =
-\nabla (b\,  \rho ) - \lambda |\Delta |^{\mu /2}\rho $. All notations are consistent with those of Refs.~\cite{gs,gar,olk}. To deal with the Cauchy driver, we set $\mu =1$.

Cauchy pdf  \eqref{ca2}  is a stationary  pdf of the Langevin-driven  fractional Fokker-Planck equation, with
 noise intensity $\lambda =1$
\begin{equation}
\partial _t\rho =
-\nabla (b\,  \rho ) - |\Delta |^{1/2}\rho.
\end{equation}
Since the process  is by construction  relaxing to  an invariant pdf  $\rho _{*}(x)$,   the  drift function
is defined as follows
\begin{equation}
b(x) = b_{jump}(x)  = -{\frac{1}{\rho _*(x)}} \int (|\nabla |\rho _*)(x)\, dx.
\end{equation}
Accordingly, by selecting the pdf \eqref{ca2}, to which the OUC process actually relaxes, we identify the
corresponding  drift function  as
  $b_{1,jump}(x)=-x$.

\subsubsection{Semigroup dynamics with Cauchy driver}

In Ref.~\cite{gs} we have investigated in some detail the  Cauchy-Schr\"{o}dinger semigroup-driven random  motion scenario
 (e.g. semigroup dynamics with  Cauchy driver). In this theoretical framework, the major ingredient of the formalism is the
(pseudo-differential, self-adjoint in a suitable Hilbert space) Hamiltonian operator  $\hat{H}_{\mu }  \doteq   \lambda
|\Delta |^{\mu /2} +  {\cal{V}}$, that  gives rise (presently, we set $\mu =1= \lambda $) to a fractional
analog of the familiar generalized diffusion equation, c.f. \cite{gar},
 \begin{equation}\label{pseudo}
 \partial _t\Psi = -  |\Delta |^{1/2} \Psi
- {\cal{V}} \Psi \, .
\end{equation}
The dynamics of the  related pdf
 \begin{equation}
 \rho (x,t) = \Psi (x,t) \rho _*^{1/2}(x)
 \end{equation}
  is fully determined by Eq.~\eqref{pseudo}, provided $\rho _*$ stands for an a priori given terminal (asymptotic) probability
   density of the involved stochastic process.

The   effective   potential ${\cal{V}}(x)$  for  the semigroup-driven jump-type  process
reads
\begin{equation}
{\cal{V}}(x)= - {\frac{|\nabla |\rho ^{1/2}_*}{\rho ^{1/2}_*}} \, .
\end{equation}
Presently, we identify $\rho _*$ with  $\rho_{*1}$. A functional form of the corresponding ${\cal{V}}(x)$
has been found  in Ref.~\cite{gs}:
\begin{equation}\label{ca4}
{\cal V}_1(x)=\frac{1}{\pi}\left[-\frac{2}{\sqrt{a(x)}}+\frac{x}{a(x)}
\ln\frac{\sqrt{a(x)}+x}{\sqrt{a(x)}-x}\right].
\end{equation}
where $a(x)=1+x^2$, c.f. also Fig.~1  in   Ref.~\cite{gs}.

 The temporal behavior of the Langevin-driven and semigroup driven process with  $\rho_{*1}(x)$ as the asymptotic pdf, has been visualized earlier  in Figs 1 and 2 of Ref. \cite{gs1}.
  Clearly,  the semigroup  potential ${\cal V}_1(x)$ is completely divorced from the harmonic potential
   $V(x)= x^2/2 \rightarrow -\nabla V = -x =b_{1,jump}(x)$, appropriate  for the Langevin-driven  OUC process.
   We observe that  $\rho_{*1}(x)$ cannot be rewritten in the Gibbs form  $(1/Z)\, \exp[- V(x)]$, in conformity with arguments of Ref.~\cite{klafter}.

\subsubsection{Wiener driver: diffusion-type process}

The previous two dynamical patterns of behavior seem to be natural in connection with the Cauchy density, since both refer to the Cauchy noise
(e.g. driver)   and the Cauchy operator  $|\nabla | =|\Delta |^{1/2}$  as the generator of random motion.
Therefore, it seems to be far from  being obvious that we can  address the previous stochastic interpolation issue
 in terms of   diffusion-type  processes, hence with the Wiener driver in action.

 For a  diffusion process  that is  governed by the standard  Fokker-Planck equation
\begin{equation}
 \partial _t\rho = {\frac{1}{2}}\Delta \rho - \nabla \, ( b \cdot
\rho )\,
\end{equation}
with a certain $\rho _*(x)$ as its stationary solution, at least on formal grounds  we  can identify the drift of the process $b(x)$ as
\begin{equation}
b = b_{diff}= {\frac{1}{2}} \nabla  \ln \rho _*  \, .
\end{equation}
Consequently, upon assuming that $b_{diff}(x)=-\nabla V_*$, we have   $  \rho _*(x) = (1/Z)\,  \exp[ - V_*(x)/k_BT]$,
where $1/Z$ is a normalization constant. This   has   the  familiar   Gibbs-Boltzmann  form, although
 $V_*(x)$ surely has nothing in common with $V(x)$ appropriate for the Cauchy-Langevin dynamics.

Indeed, under  the diffusion (i.e. Langevin-driven) process premises,
\begin{equation}
\dot{x} = b(x,t) +A(t)
\end{equation}
 where $\langle A(s)\rangle =0$, $\langle
A(s)A(s')\rangle =  \delta (s-s')$, a demand that  an asymptotic pdf is  in the  Cauchy form Eq.~\eqref{ca2},
enforces  the drift function
\begin{equation}\label{ca5}
b_{1,diff}(x)=-\frac{2x}{1+x^2}\, ,
\end{equation}
which is a gradient field with  the  potential $V_{*1}(x)= \ln (1+x^2)$.
These expressions have  been substituted to the corresponding numerical routines  to obtain a computer-assisted
picture of the time evolution $\rho_1(x,t)$ and get confirmed that such  diffusion processes  are consistent.

\begin{figure}[!h]
\begin{center}
\includegraphics [width=0.9\columnwidth]{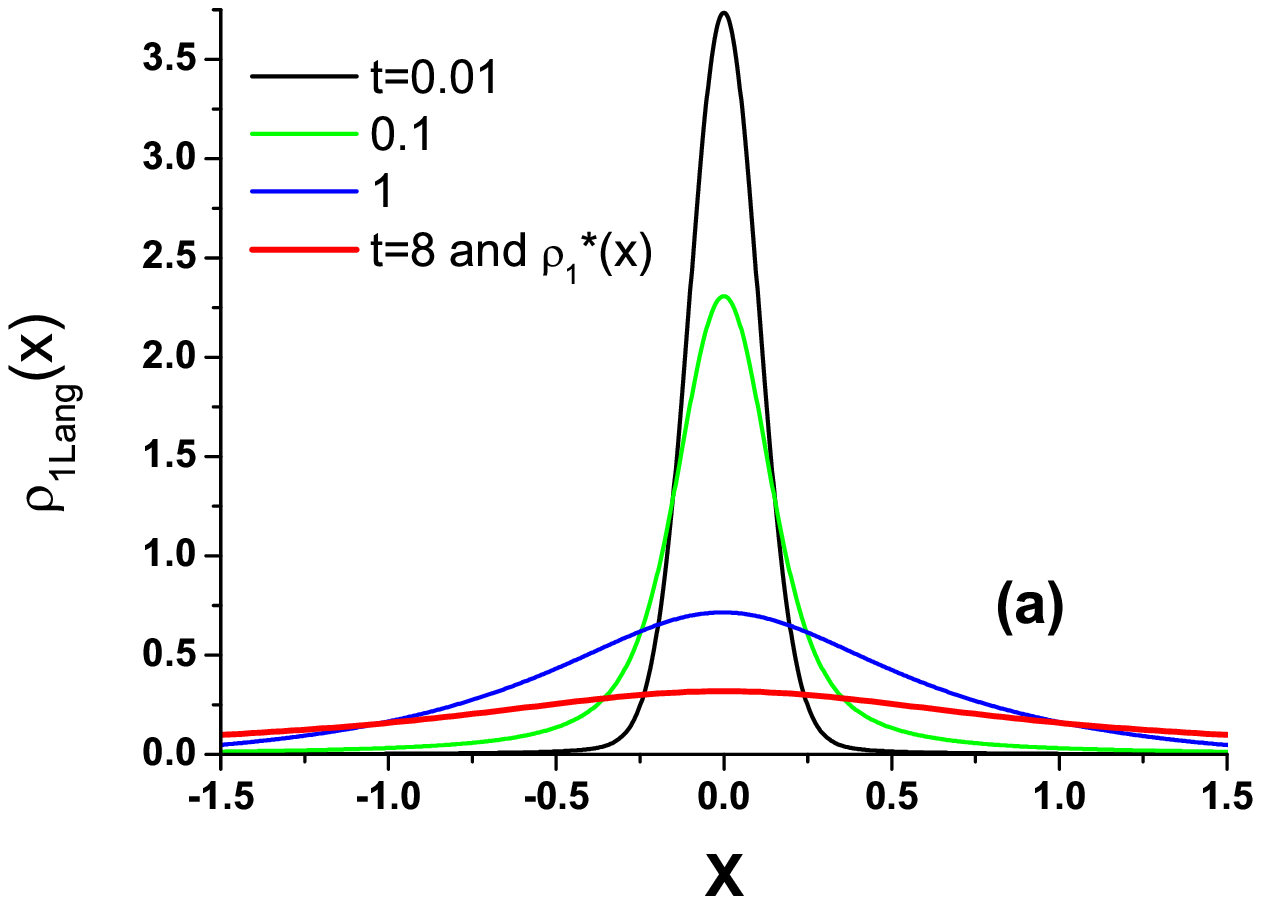}
\includegraphics [width=0.9\columnwidth]{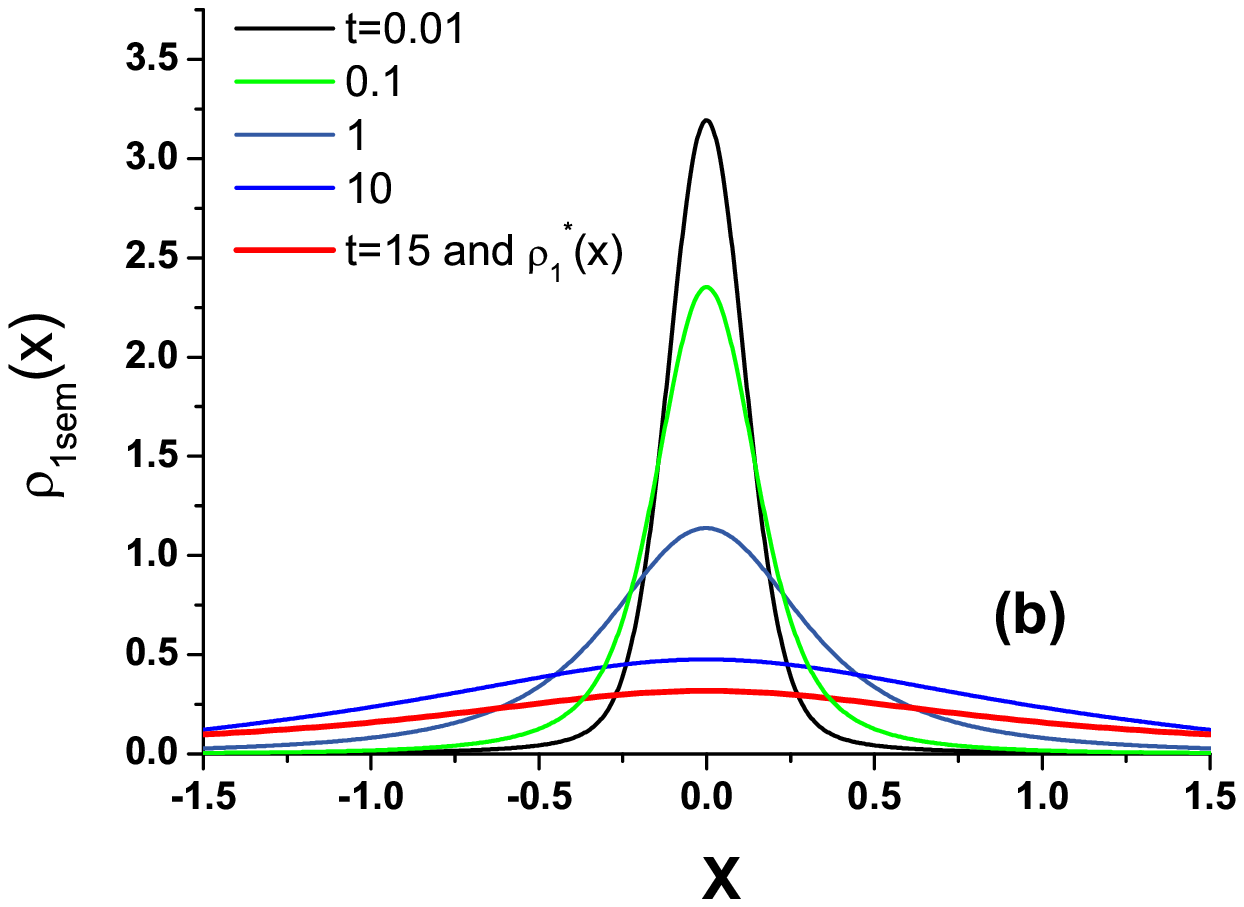}
\includegraphics [width=0.9\columnwidth]{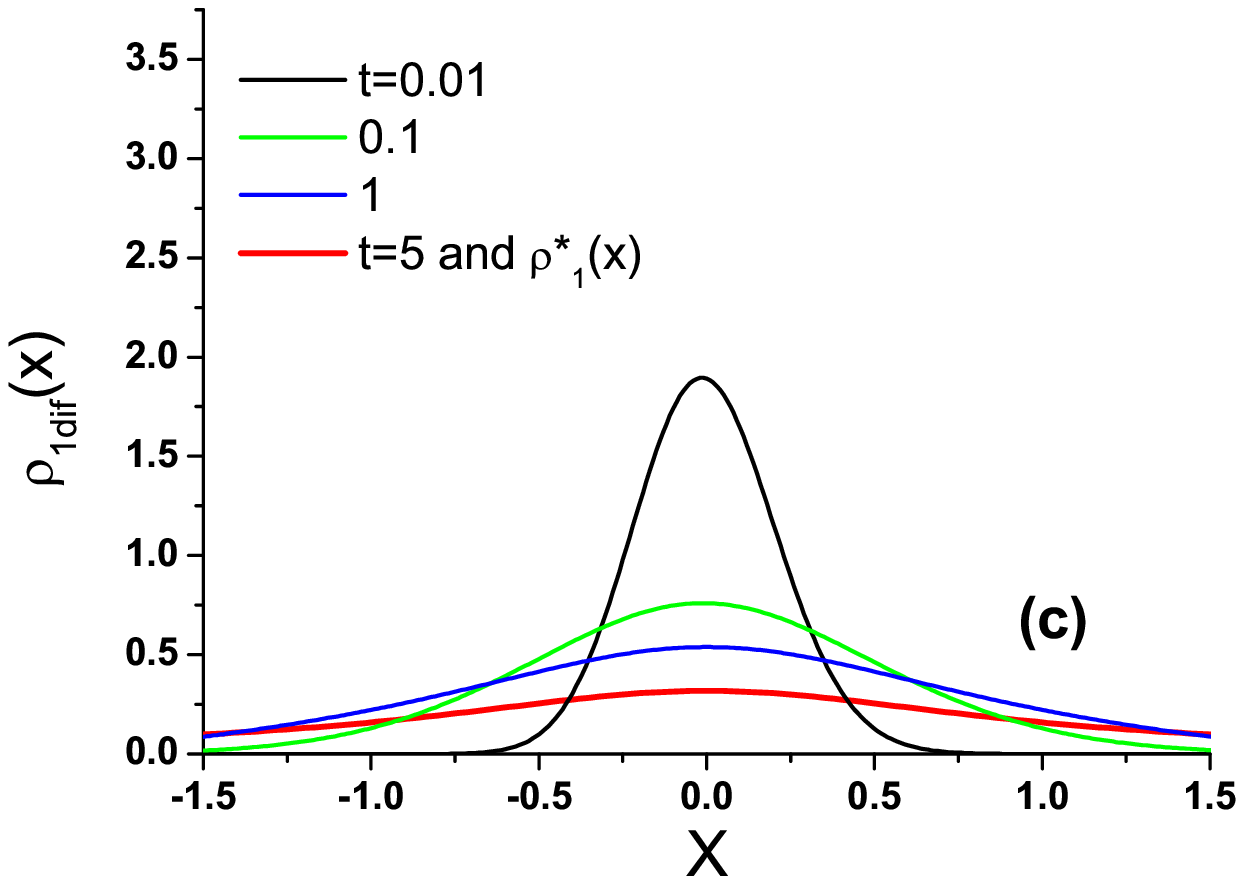}
\end{center}
\caption{Time evolution of pdf's $\rho(x,t)$ for the Cauchy-Langevin
dynamics (panel (a)), Cauchy-semigroup-induced evolution (panel (b))
and the  Wiener-Langevin  process (panel (c)). The common
target pdf is the Cauchy density, while the initial $t=0$ pdf is set
to be  a Gaussian with height 25 and half-width $\sim 10^{-3}$. The first depicted stage of evolution
corresponds to  $t=0.01$.  The time rate hierarchy seems to be set: diffusion being fastest, next L\'{e}vy-Langevin  and semigroup-driven evolutions being slower than previous two. However the outcome is not universal, as our subsequent discussion will  show.} \label{fig:ca1}
\end{figure}

A comparison of the above  three different (Cauchy-Langevin, Cauchy semigroup and Wiener-Langevin)  stochastic interpolation scenarios,
 all started form identical initial data (Dirac delta approximation by a Gaussian) data and all terminating (in the large time asymptotic)
 at common  for all  Cauchy pdf, is provided in Fig.~\ref{fig:ca1}.

The diffusion-type dynamics is faster than that of the Langevin-driven  jump-type
process.  On the other hand, to achieve the target pdf, the semigroup dynamics needs almost twice a time ($t=15$) as that ($t=8$)
 needed for Cauchy - Langevin relaxation.
In Fig.~\ref{fig:ca1}, we have indicated  time instants at which the simulated pdf, within the resolution quality of the figures, cannot be
 distinguished from the target pdf $\rho _{*1}$.

 The above   relaxation pattern is generic for  Cauchy family of pdfs.
 However, the  resulting  "speed" hierarchy is not generic for  stochastic processes at hand.
 Below we shall demonstrate that  time rates of relaxation  for our three exemplary  processes
 (i.e. jump-type Langevin, jump-type semigroup, diffusion-type)  do not follow any definite  hierarchy.

\subsection{Second order target, $\rho_{*2}$.}

With the basic notation specified, we can proceed in quick steps. Let us consider
the $n=2$  pdf of the Cauchy hierarchy
\begin{equation}\label{ca6}
\rho_{*2}=\frac{2}{\pi(1+x^2)^2}.
\end{equation}
If regarded as an invariant density of the Cauchy-Langevin stochastic process, this pdf gives rise to the following drift function
\begin{equation}\label{ca7}
b_{2,jump}(x)=-\frac x8(x^2-3)\, .
\end{equation}

The semigroup-driven process,  relaxing to  $\rho_{*2}(x)$  in the large time asymptotics,  is determined by specifying the semigroup potential, \cite{gs} c.f. Fig.~1 there-in:
\begin{equation}\label{ca8}
{\cal V}_2(x)=\frac{x^2-1}{x^2+1}.
\end{equation}
The affiliated diffusion process (Wiener driver), with the very  same asymptotic pdf  has the  drift function
\begin{equation}\label{ca9}
b_{2,diff}(x)\equiv 2b_{1,diff}(x) = - {\frac{4x}{1+x^2}}.
\end{equation}
At this point, we note a   recursive formula $b_{n,diff}(x)\equiv nb_{1,diff}(x)$ and $V_{*n}(x) = n\, \ln (1+x^2)$.

Qualitatively,  the temporal behavior of the three involved stochastic processes mimics the pattern of $n=1$, according to Fig.~\ref{fig:ca1},
except that with the growth of $n$ the speed with which the processes relax  increases.

\subsection{Third order target, $\rho_{*3}$.}

The third, n=3 member,  of the  pdfs  family reads
\begin{equation}\label{ca11}
\rho_{*3}=\frac{8}{3\pi(1+x^2)^3}.
\end{equation}
It  is an invariant probability density of the Cauchy-Langevin process with the  drift function
\begin{equation}\label{ca12}
b_{3,jump}(x)=-\frac {x}{16}(15+10x^2+3x^4)\, .
\end{equation}
An effective  potential for the Cauchy-Schr\"{o}dinger dynamics can be evaluated by means of the Cauchy principal value integrals. Following methods of  Ref.~\cite{gs}  we  have arrived at its  explicit functional  form
\begin{equation}\label{ca13}
{\cal V}_{3}(x)=\frac{1}{\pi}\left[\frac{2(x^2-2)}{\sqrt{a(x)}}+\frac{3x}{a(x)}
\ln\frac{\sqrt{a(x)}+x}{\sqrt{a(x)}-x}\right].
\end{equation}
where $ a(x)=1+x^2$.
 Like the previous two semigroup potentials, that were bounded from below and above,
  the present one,  although unbounded from above, still  fulfills minimal  requirements  set in the general theoretical framework
  of   Ref.~\cite{olk}. It is bounded from below  by ${\cal V}_3(0)=-4/\pi$ and
   for $|x|\to \infty$  we have  ${\cal V}_3(x)\approx 2|x|/\pi$.

The diffusion process with $\rho_{*3}(x)$ as its asymptotic target is characterized by the drift function $ b_{3,diff}(x)= - 6x/(1+x^2)$.

\subsection{Fourth order target, $\rho_{*4}$.}
Although we know that nothing illuminating might  happen in the
qualitative picture of the time  evolution, if compared to the
previous cases, for completeness we provide  drift functions and  the semigroup potential, that can be derived if n=4
pdf is to be asymptotic target of, respectively: Cauchy-Langevin, Cauchy semigroup and diffusive motions:
\begin{equation}\label{ca14}
\rho_{*4}=\frac{16}{5\pi(1+x^2)^4}.
\end{equation}
The   Cauchy-Langevin drift presently  reads
\begin{equation}\label{ca15}
b_{4,jump}(x)=-\frac {x}{16}(35+35x^2+21x^4+5x^6)\,  ,
\end{equation}
the Cauchy-Schr\"{o}dinger semigroup potential appears in the form
\begin{equation}\label{ca16}
{\cal V}_4(x)=\frac{x^4+6x^2-3}{2(x^2+1)}.
\end{equation}
while   the  drift of an affiliated diffusion process   is  $b_{4,diff}(x)= -8x/(1+x^2)$.

\subsection{Tail (large $|x|$)  regularities}

It is easily  seen from the  above equations that as $|x|\to \infty$, we encounter some
regularities. Namely,  pdfs in the Cauchy hierarchy for large $|x|$   behave like
\begin{equation}
 \rho _{*n}(x) \approx  {\frac{1}{x^{2n}}}\, .
 \end{equation}
 Cauchy-Langevin drifts  have large $|x|$ asymptotics of the form
\begin{equation}\label{cas1}
b_{n,jump}(x\to \infty)\approx x^{2n-1}.
\end{equation}
 Cauchy-Schr\"{o}dinger semigroup potentials behave like
\begin{eqnarray}
&&{\cal V}_1(|x|\to \infty)\approx \frac{2\ln|x|}{|x|}\sim \frac 1{|x|},\label{v1}\\
&&{\cal V}_2(|x|\to \infty)\approx 1\equiv |x|^0,\label{v2}\\
&&{\cal V}_3(|x|\to \infty)\approx |x|,\label{v3}\\
&&{\cal V}_4(|x|\to \infty)\approx |x|^2,\label{v4}
\end{eqnarray}
so that for arbitrary $n$
\begin{equation}\label{cas2}
{\cal V}_n(|x|\to \infty)\approx |x|^{n-2}.
\end{equation}

The forward drifts of a diffusion - type processes, for which  Cauchy family pdfs  \eqref{ca1} are
their respective asymptotic targets, read
\begin{equation} b_{n,diff}(x)
\approx -{\frac{2n}{x}}.
\end{equation}

\begin{figure*}
   \subfigure[\hspace*{0.5mm}Half-width and $<X^2>$ (divergent) for $\rho_{*1}$ ]
   {\includegraphics*
  [width = 85mm]{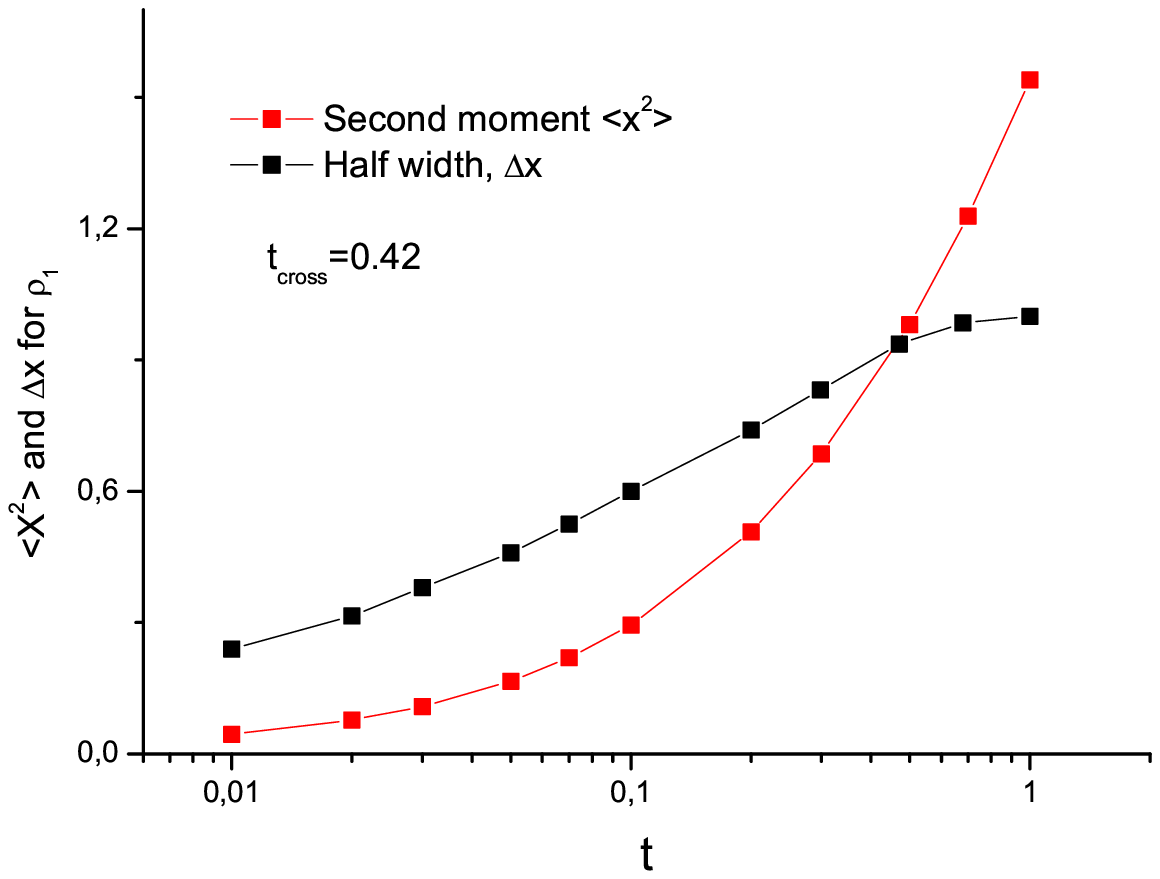}
    \label{ex1}}
    \subfigure[\hspace*{0.5mm}$<X^2>$ and $<X^4>$ (divergent) for $\rho_{*2}$]{\includegraphics*
   [width = 80mm]{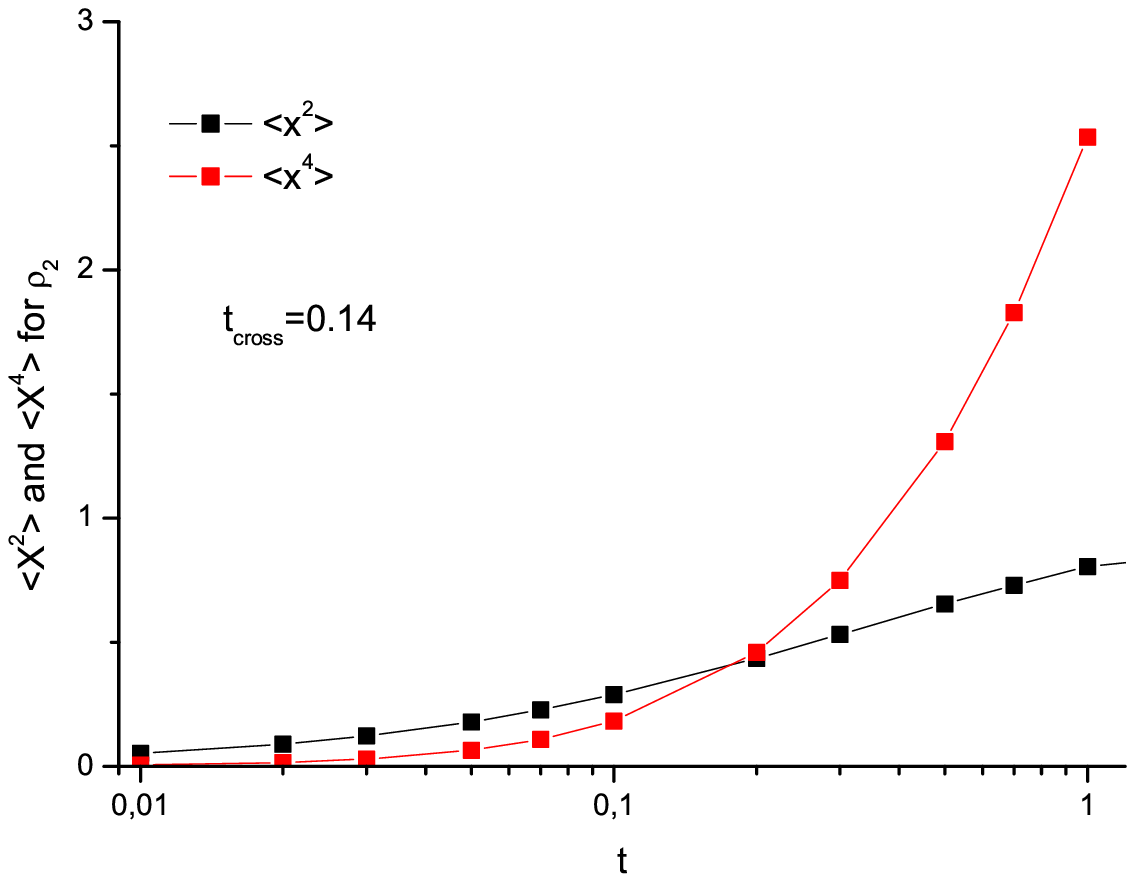}
    \label{si1}}
    \subfigure[\hspace*{0.5mm}$<X^4>$ and $<X^6>$ (divergent) for $\rho_{*3}$]{\includegraphics*
  [width = 80mm]{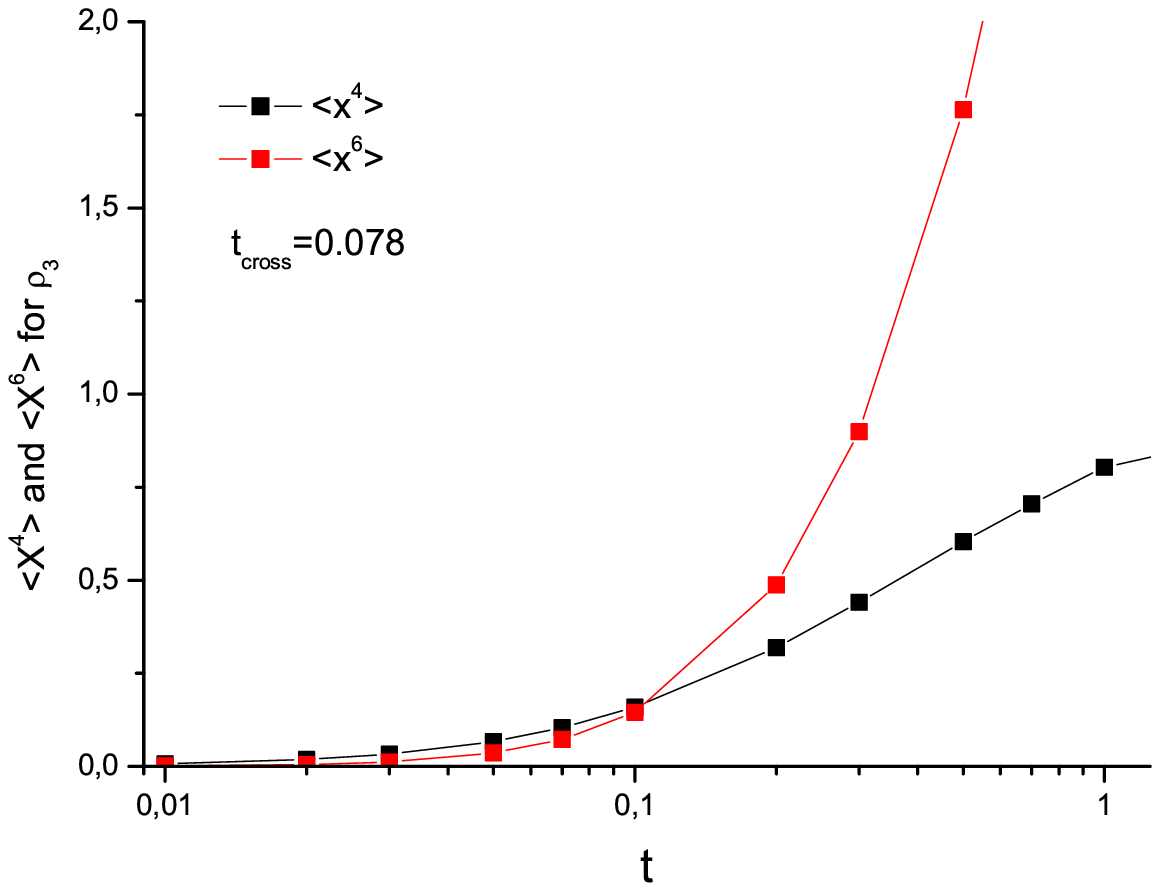}
    \label{ex2}}
    \subfigure[\hspace*{0.5mm}$<X^6>$ and $<X^8>$ (divergent) for $\rho_{*4}$]{\includegraphics*
   [width = 80mm]{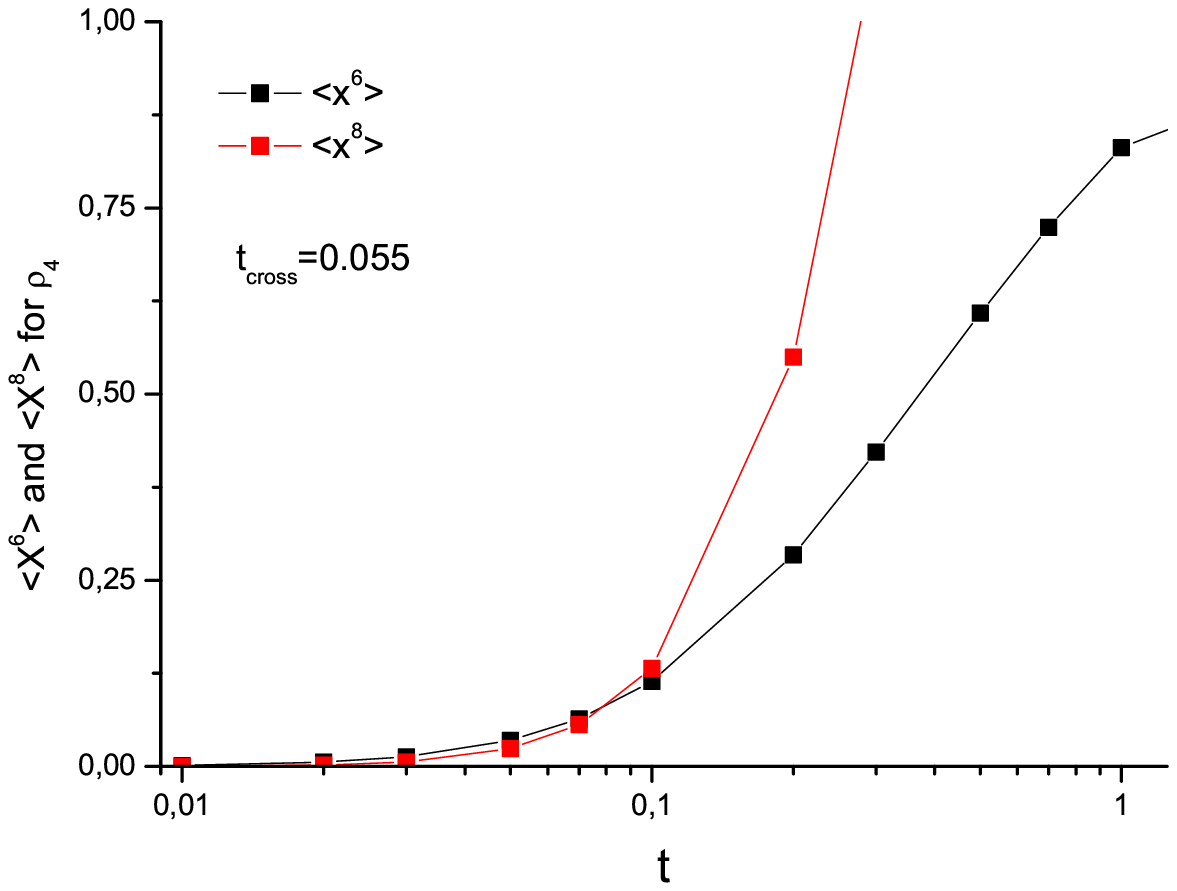}
    \label{si2}}

    \caption{The time evolution of the last convergent and first divergent moment for pdfs $\rho_{*1}$ - $\rho_{*4}$, shown on panels (a) - (d) respectively. Crossover time instants are shown on the corresponding panels. } \label{fig:ca2}
\end{figure*}

\subsection{A cross-over between exponential and power law behavior}

We have mentioned before that the Cauchy hierarchy  is ordered with respect to  an overall  number of moments in existence,
 of the probability distribution. Namely, for $n=1$ even the first moment does not exist.
 In principle (if Cauchy principal value of the corresponding integral is considered) we can
admit  the existence of the first moment. But still, the second moment does not exist.

For $n=2$, we have first three moments, the  third one exists conditionally if the above Cauchy principal value is considered.
Ultimately,  each  $\rho _{*n}$ has exactly $2n-1$ moments for each  $n\geq 1$.

On the other hand, in our numerical procedures,
the nonlinear (in a sense that corresponding drift function is not linear) diffusion - type motion is started from a very narrow
Gaussian. As  indicated  in the caption to  Fig.~\ref{fig:ca1}, its  height (maximum) is  $\sim 25$  while  half-width $\sim 10^{-3}$. This pdf certainly admits all moments.

   In the course of the   diffusion - type  evolution  with  a target pdf belonging to the Cauchy family, quite apart from
   incompatible microscopic mechanisms (jumps vs continuous paths)  there must arise a definite cross-over
  from  Gaussian to  Cauchy hierarchy  behavior. The latter hierarchy exhibits the power law asymptotics to be set
against Gaussian (stronger then exponential) decay. Therefore higher moments of the initial  pdf, while evolving  diffusively,
must consecutively disappear in the course of time, if any Cauchy density is to be a target pdf.

To visualize this intriguing crossover between  Gaussian and Cauchy family behaviors,
 we have numerically compared the   time  development of largest finite moment for a given $n$ (which is  $(2n-2)$-th
moment of Cauchy family \eqref{ca1} pdfs) with that of
a smallest nonexistent moment (which is $2n$-th moment of the same family). This evolution has been generated by
Cauchy - Langevin  driven  diffusion-type $\rho (x,t)$ beginning (at $t=0$) at above very narrow Gaussian and ending (at $t \to \infty$)
at $\rho _{*n}$ \eqref{ca1}. This has been done for $n=2-4$ and is reported in Fig.~\ref{fig:ca2}.

The special case $n=1$ has been considered separately. Namely, following our earlier results, Ref. \cite{gs1},  we consider the half-width at half-maximum of function \eqref{ca2} instead of  the above largest finite moment. Then, similar to the above $n>1$ case, we compare the time evolution of this quantity in the course of the  OUC process,
 with that of the second moment for a diffusion-type probability distribution $\rho (x,t)$ having $\rho _{*1}(x)$
\eqref{ca2} as the target (asymptotic) pdf, see Fig. \ref{fig:ca2}a.

\subsection{Interpolation between Cauchy and Gaussian
distributions}

Since, within the Cauchy hierarchy, the number of moments increases with $n$,
we  can justifiably ask for its link with the Gaussian pdf. At least, for large $n$.

Let us rewrite Eq. \eqref{cf} in the form
\begin{equation}\label{tf1}
    \rho_\alpha (x)= \frac{A_\alpha}{(1+x^2)^\alpha}\equiv \exp (\ln \rho_\alpha)\, ,
\end{equation}
where $\ln \rho_\alpha=\ln A_\alpha-\alpha \ln(1+x^2)$.

We observe that  the limiting procedure  $\alpha \to \infty$  enforces strong localization properties
of the pdf so that only a small vicinity of $x=0$ matters. Indeed, at large $\alpha$, the pdf tails
become very steep and do not play a decisive role. This motivates  the following series expansions
\begin{eqnarray}
\ln (1+y)&=&y-\frac{y^2}{2}+\frac{y^3}{3}-\frac{y^4}{4}+...,\ y=x^2,  \label{lny1}\\
A_\alpha (\alpha \to
\infty)&=&\sqrt{\frac{\alpha}{\pi}}-\frac{3}{8\sqrt{\pi
\alpha}}-\frac{7}{128\alpha\sqrt{\pi \alpha}}-..\  . \label{lny2}\, .
\end{eqnarray}
Accordingly:
\begin{equation}\label{tf2}
    \rho_\alpha (x,\alpha \gg 1)\approx\left[\sqrt{\frac{\alpha}{\pi}}-\frac{3}{8\sqrt{\pi \alpha}}-\frac{7}{128\alpha\sqrt{\pi \alpha}}-..\right]\times
\end{equation}
$$
\exp\left(-\alpha x^2 +\alpha \frac{x^4}{2}-...\right),
$$
It is obvious that leading terms (in the large $\alpha$ regime) coincide with a perfect Gaussian
\begin{equation}\label{fin}
    \rho_\alpha (x,\alpha \gg 1) \approx \sqrt{\frac{\alpha}{\pi}}\exp(-\alpha x^2).
\end{equation}
We can easily check that function \eqref{fin} is normalized.
Moreover,  Eq. \eqref{tf2}  sets  a "bridge"  between  the  Gaussian and  large $\alpha $ members of the  Cauchy
family.   The  account  of  a  sufficiently large number  of terms in the  series
\eqref{tf2}, permits to control an accuracy  of the   "Gaussian approximation" of Cauchy family pdfs with large $\alpha$.
Ultimately, since $\alpha \to \infty $  stands for a sequential approximation of the Dirac delta functional, our limiting procedure within the Cauchy family provides another  sequential approximation of the  Dirac delta.

\begin{figure*}
\begin{center}
\includegraphics [width=0.7\columnwidth]{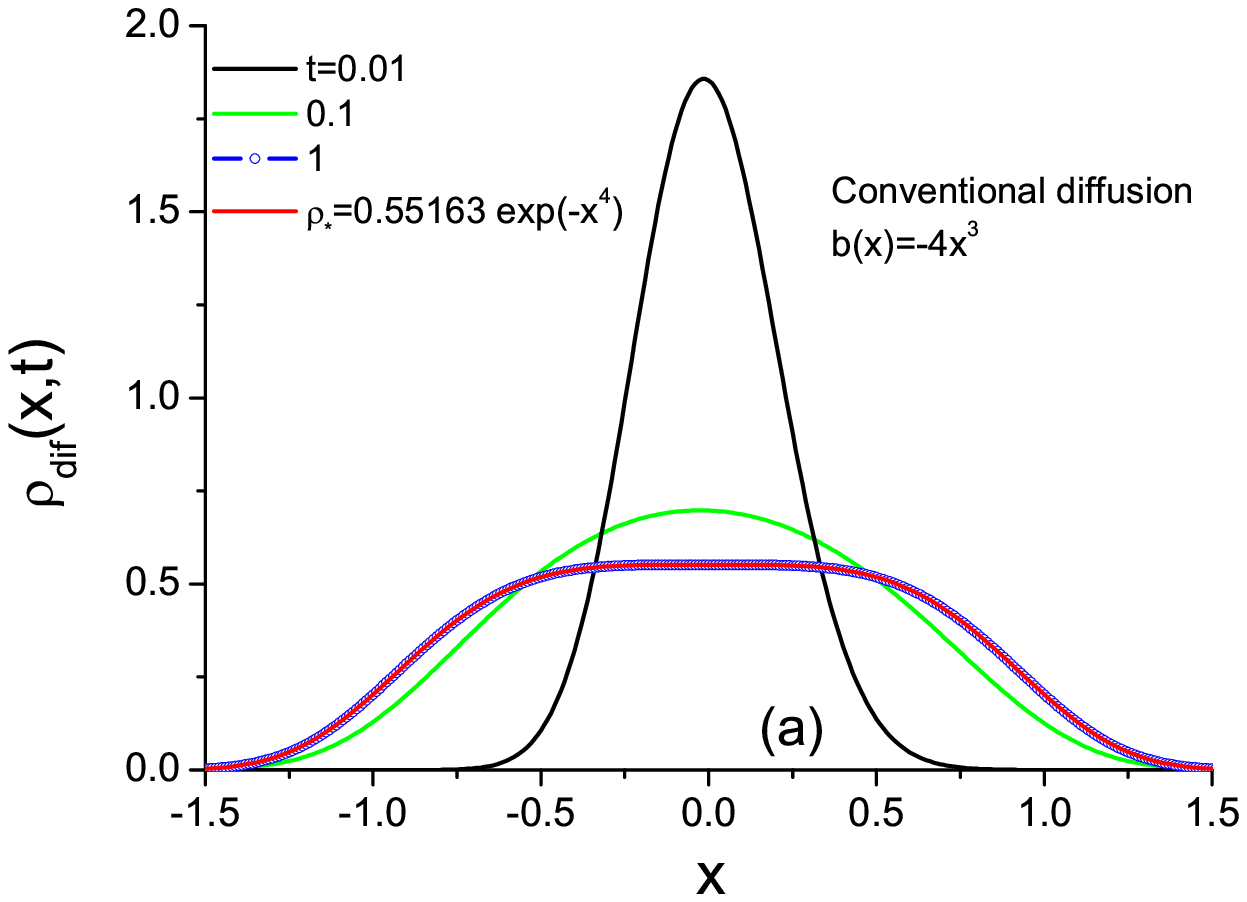}
\includegraphics [width=0.7\columnwidth]{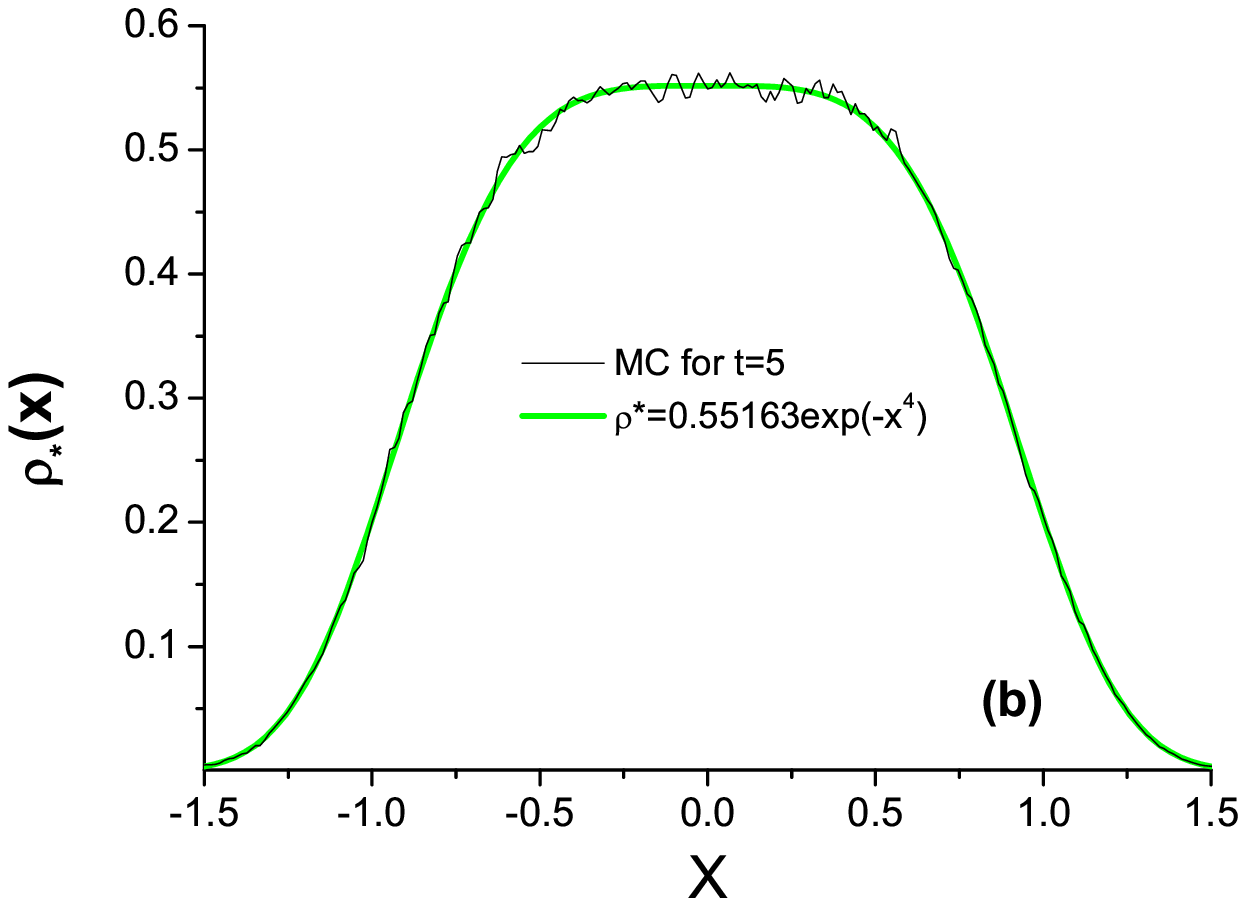}
\includegraphics [width=0.65\columnwidth]{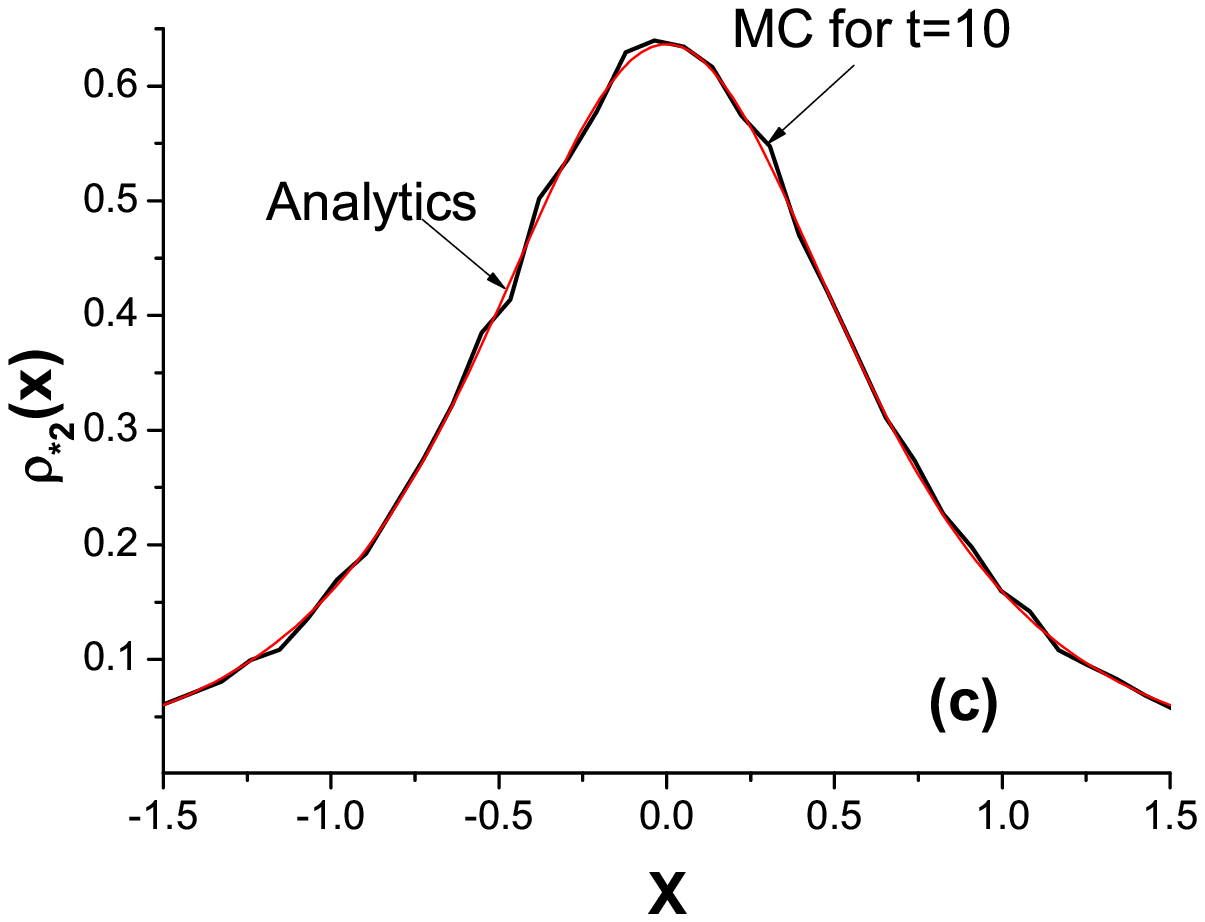}
\end{center}
\caption{Time evolution of a  diffusion process with drift $b(x)=-4x^3$. Panel (a) shows the pdf dynamics, obtained from FP equation. Panel (b) reports the comparison of path-wise MC simulation for $t=5$ with analytical result for asymptotic pdf.  Simulated pdf has been intentionally left "noisy" (i.e. created with smaller number of sample trajectories) to reveal the coincidence with analytical curve.  Panel (c) compares the MC result at time $t=10$ with the  analytical result for $\rho_{*2}(x)=(2/\pi)/(1+x^2)^2$. The  simulated pdf has been intentionally left "noisy". }
\label{fig:x4}
\end{figure*}

\section{Generic relaxation patterns}

\subsection{Regularity conditions in stochastic modeling}

\subsubsection{Jeopardies}
The  present subsection is motivated by various  regularity conditions that guarantee
the existence and uniqueness of solutions, both for  stochastic differential
equations and    Fokker-Planck  equations
(fractional being included),  see Refs. \cite{sobczyk,lefever,mackey,kampen} and \cite{chechkin,fogedby}.
 Those mathematical restrictions  do not  seem to worry  physics-oriented practitioners and we shall indicate
  why in some cases  this disregard might be justified.

A transparent  picture of   jeopardies to be met, comes within the traditional Brownian
motion  framework. There,   drifts and so-called  diffusion  functions (in our case, they reduce to diffusion coefficients)
 need to obey the   Lipschitz and growth bounds, \cite{sobczyk,mackey}.
It is easy to generate  Langevin and   Fokker-Planck equations with drifts that violate the growth condition. Then,
some care is necessary,  because the general theory tells us  that  the  stochastic process
in question may explode in a finite time, \cite{sobczyk}.

Polynomial  drift functions typically violate the  growth  condition.
We  shall analyze a possible  significance of this violation, by employing: (i) numericallly generated  solution
of the Fokker-Planck equation,  (ii) trajectory - wise  Monte Carlo (MC) simulations of pdfs,
in terms of Langevin equation-generated bunch (c.a. $10^5$) of  Brownian paths.

 We demonstrate  that, despite the violation of the  growth condition,
 no hint of a possible explosion is detected and the dynamical pattern of  behavior remains regular  up  to arbitrarily large times,
 available in our numerical simulations, both in FPE solution and in direct MC simulations.
Note, that in preceding discussion we have already invoked the FPE (both standard and fractional) with polynomial drifts.
A number of examples can also be found in Refs. \cite{dubkov0, dubkov,mackey}.

\subsubsection{Cubic drifts}

Let us  consider a cubic  function $b(x) = - 4x^3$ as an exemplary drift in the  F-P equation. This
 function surely does not obey the  growth condition \cite{sobczyk,mackey}: $|b(x)| + 1\leq C (1+ |x|)$.
 Standard arguments \cite{risken,mackey}   allow us  to  identify a stationary  pdf associated with  this drift function. It has
 an explicit   Gibbs form $\rho _* = (1/Z) \exp (- V_*)$, where $V_*=x^4$.

To get  convinced  that  it is actually an asymptotic pdf of a well defined   diffusion-type  process, we address the
 time evolution of  $\rho(x,t)$  that  is started  from "almost" Dirac
 delta pdf and  propagated in accordance with the corresponding FPE.
  The resultant  relaxation to the non-Gaussian
 equilibrium pdf $\rho _*(x) \sim  \exp(-x^4)$,  is depicted in  Fig.~\ref{fig:x4}a.

This relaxation pattern  has been directly   path-wise  confirmed by executing the MC simulation of  c.a.  $10^5$  Brownian paths,
 up to the time instant  $t=50$. This allowed us to   check the shape and stability of the  corresponding pdfs.
Fig. \ref{fig:x4}b reports a   direct  comparison between  the simulated (up to time  $t=5$ as for $t>5$ the corresponding pdfs are exactly the same)
and analytic asymptotics $\rho _*(x) =0.55163 \exp(-x^4)$.

Our simulation shows, that a perfect agreement of two different  generation methods  for a given pdf  extends to (arbitrarily)  large times.
Each of the  transitional   pdfs  shown in Fig. \ref{fig:x4}a  can  be   consistently
reproduced trajectory-wise,   i.e. by means  the  MC simulation whose record  is stored and analyzed at each  required time instant.
 In  Fig. \ref{fig:x4}b, we   have  intentionally   allowed the simulated asymptotic pdf to be  somewhat  "noisy".
 That outcome is a consequence of  not too large   number of  sample  trajectories  employed  for   the pdf  approximation.
Thus, on computer-assisted grounds, the time evolution of  pdfs generated by FPE's  with  polynomial  drifts,
appears to be  consistent.

\begin{figure*}
   \subfigure[]  {\includegraphics* [width = 61mm]{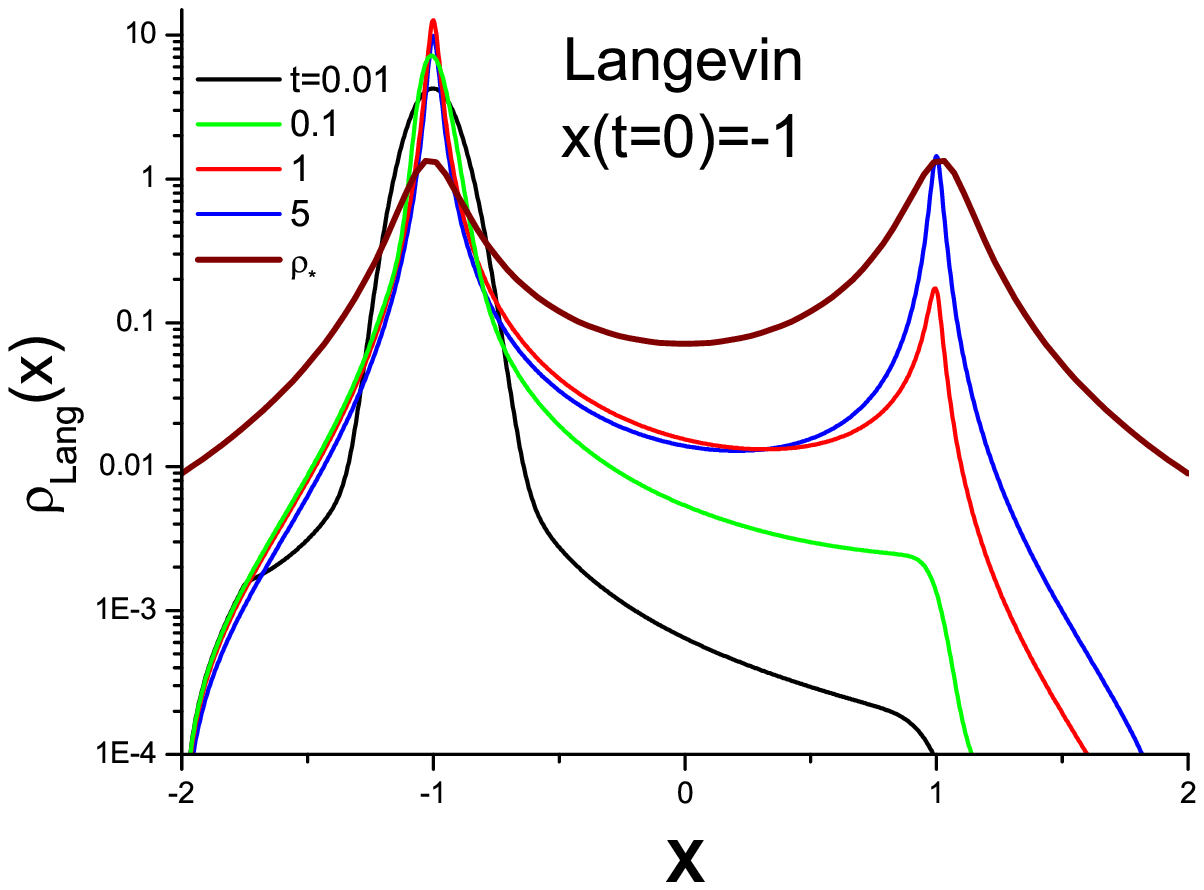}}
\hspace*{-0.5cm}
    \subfigure[]{\includegraphics* [width = 61mm]{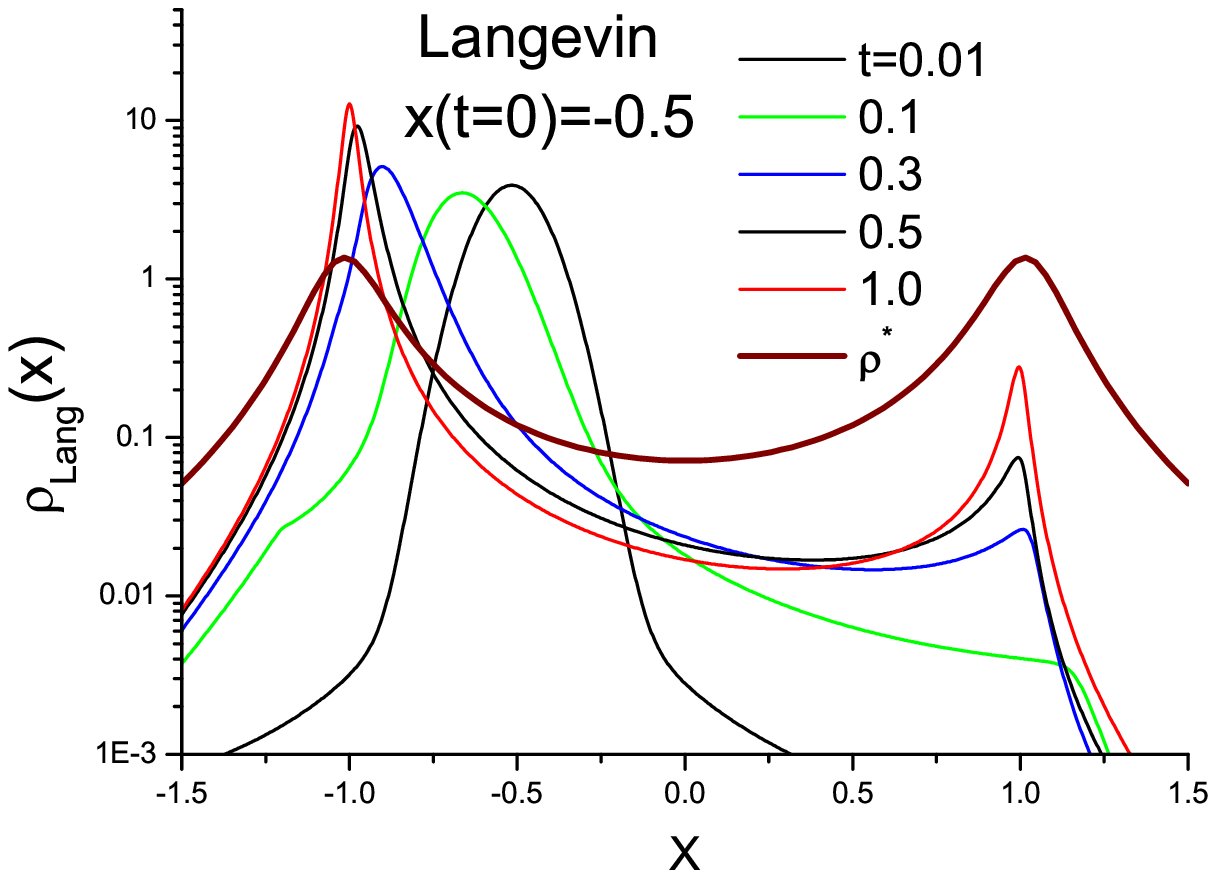}}
\hspace*{-0.5cm}
    \subfigure[]{\includegraphics* [width = 61mm]{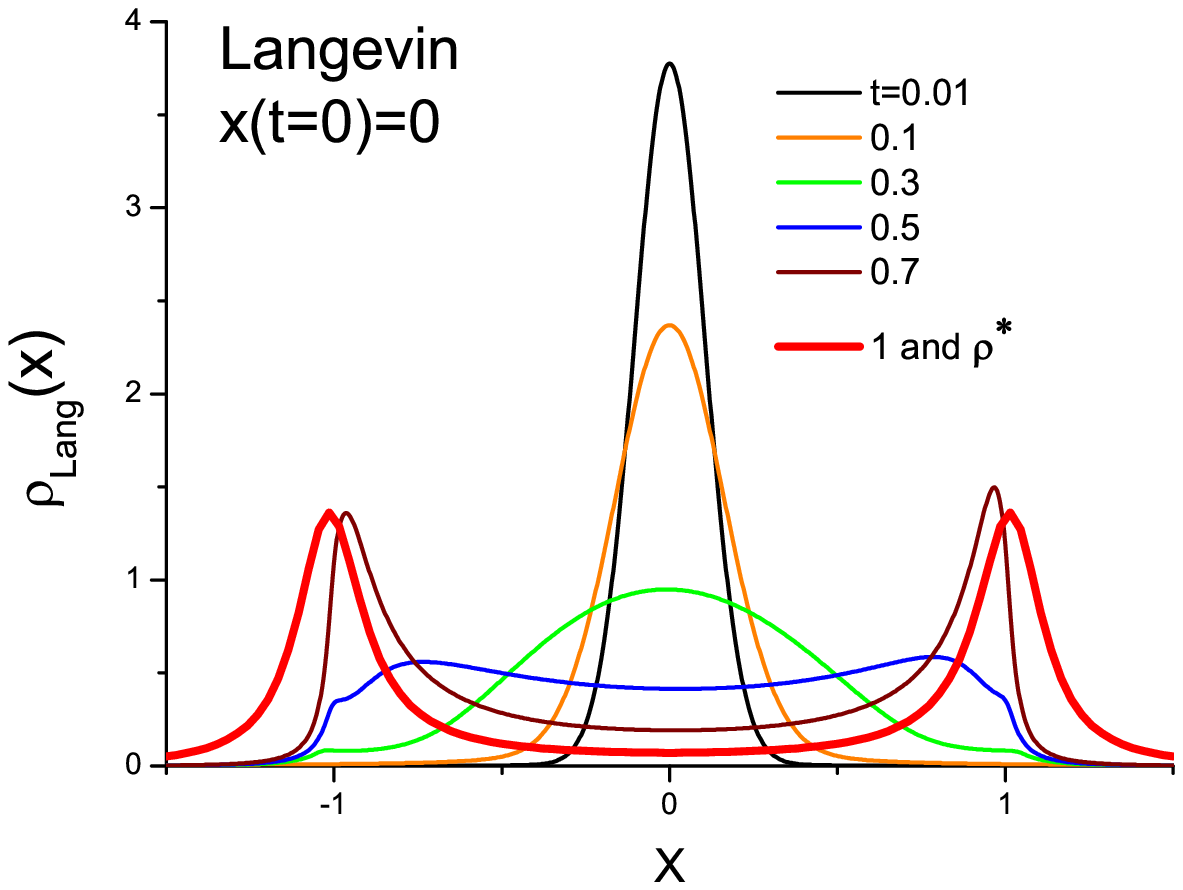}}
 \hspace*{-0.5cm}
    \subfigure[]{\includegraphics* [width = 61mm]{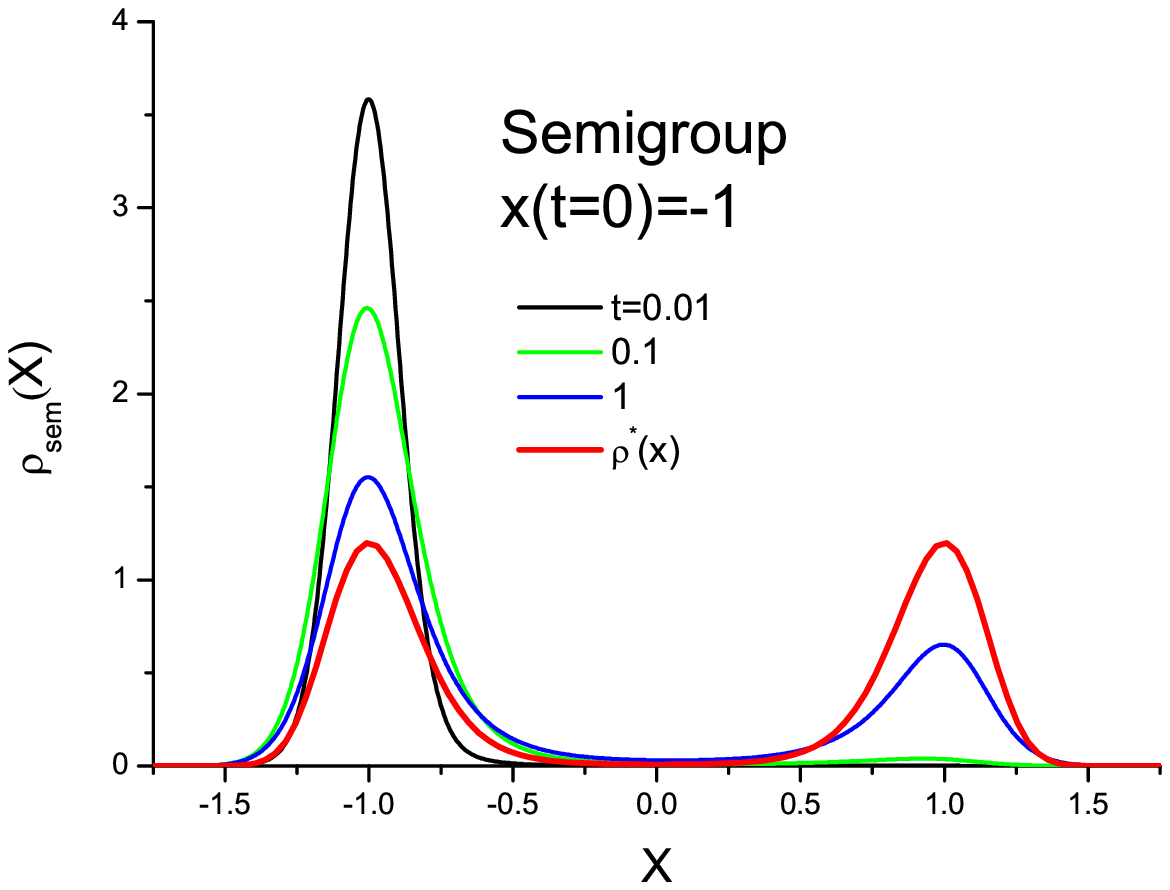}}
\hspace*{-0.5cm}
    \subfigure[]{\includegraphics* [width = 61mm]{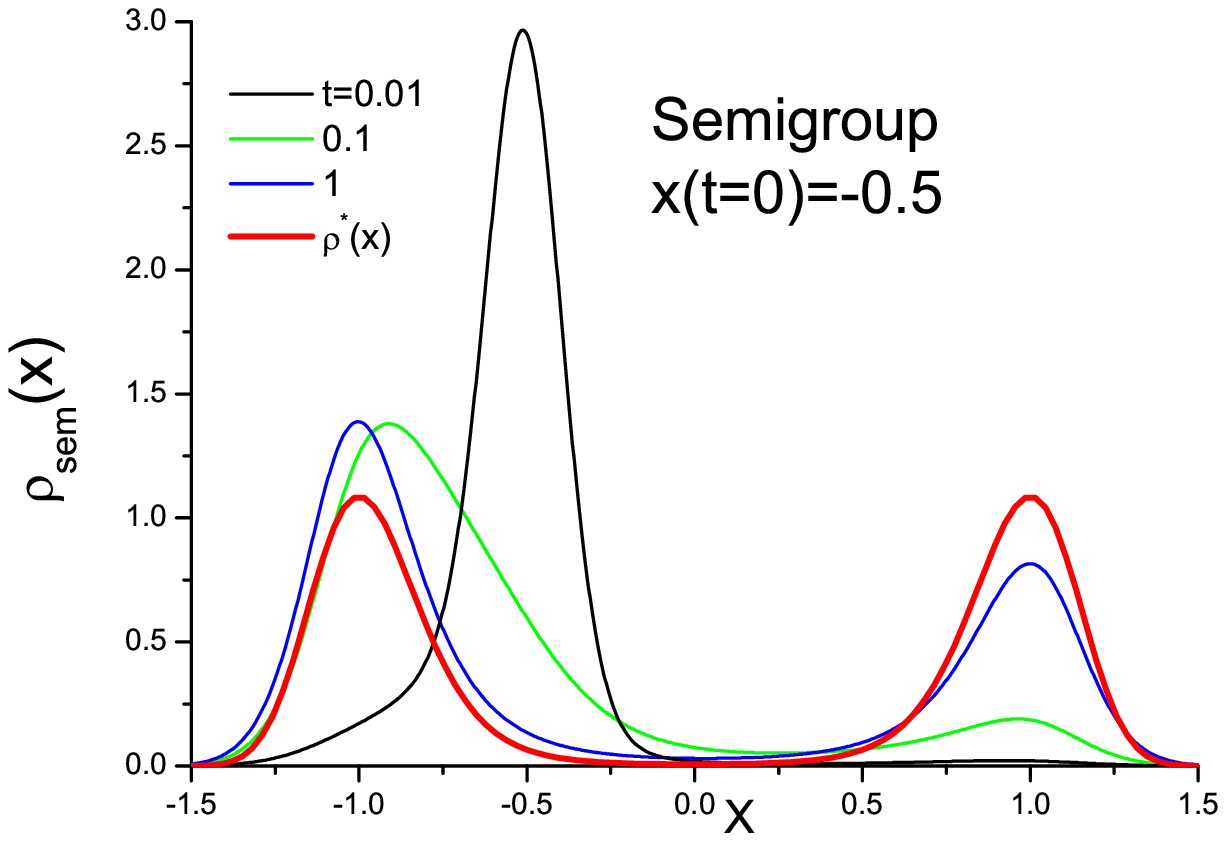}}
\hspace*{-0.5cm}
    \subfigure[]{\includegraphics*  [width = 61mm]{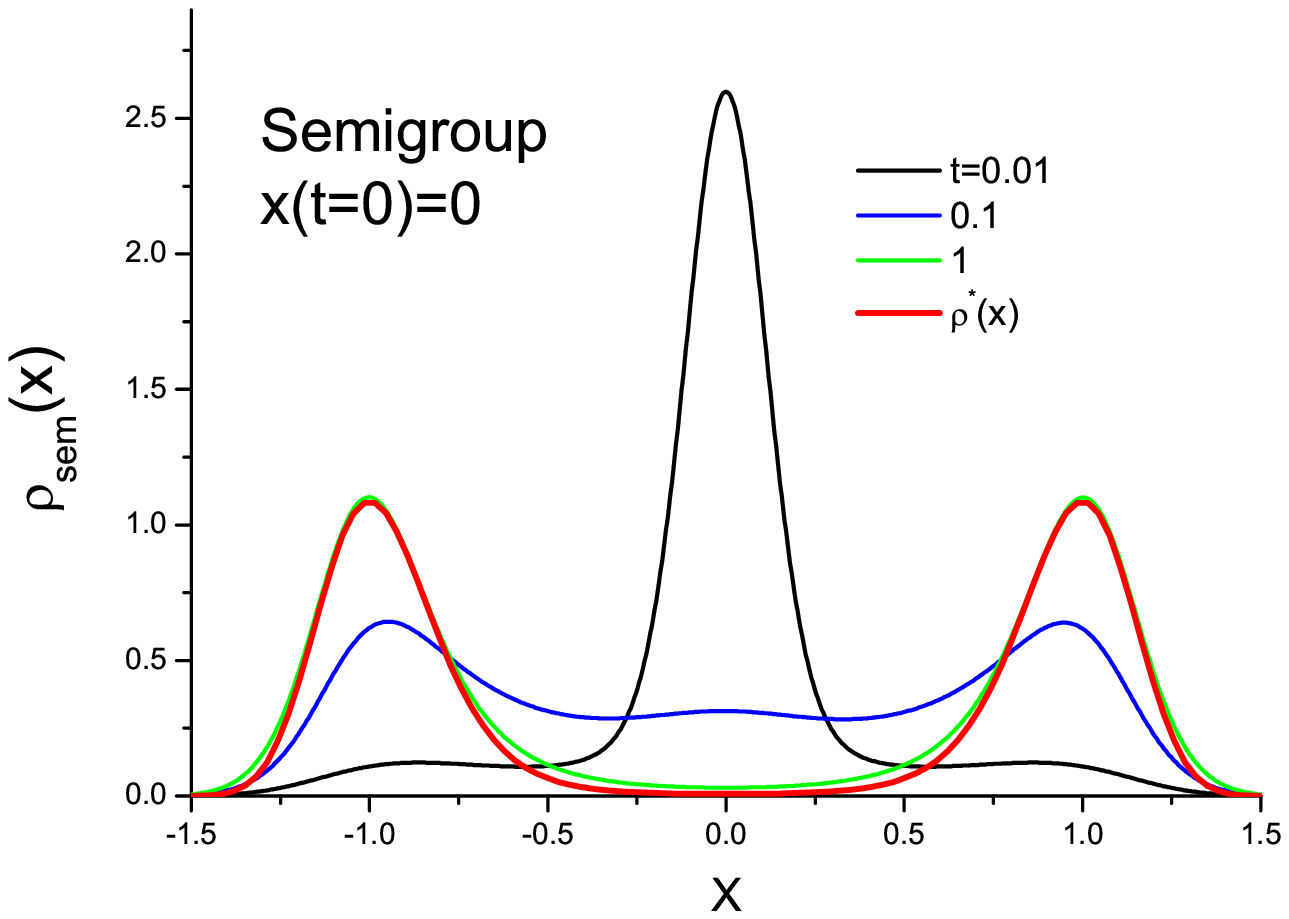}}
\hspace*{-0.5cm}
    \subfigure[]{\includegraphics* [width = 61mm]{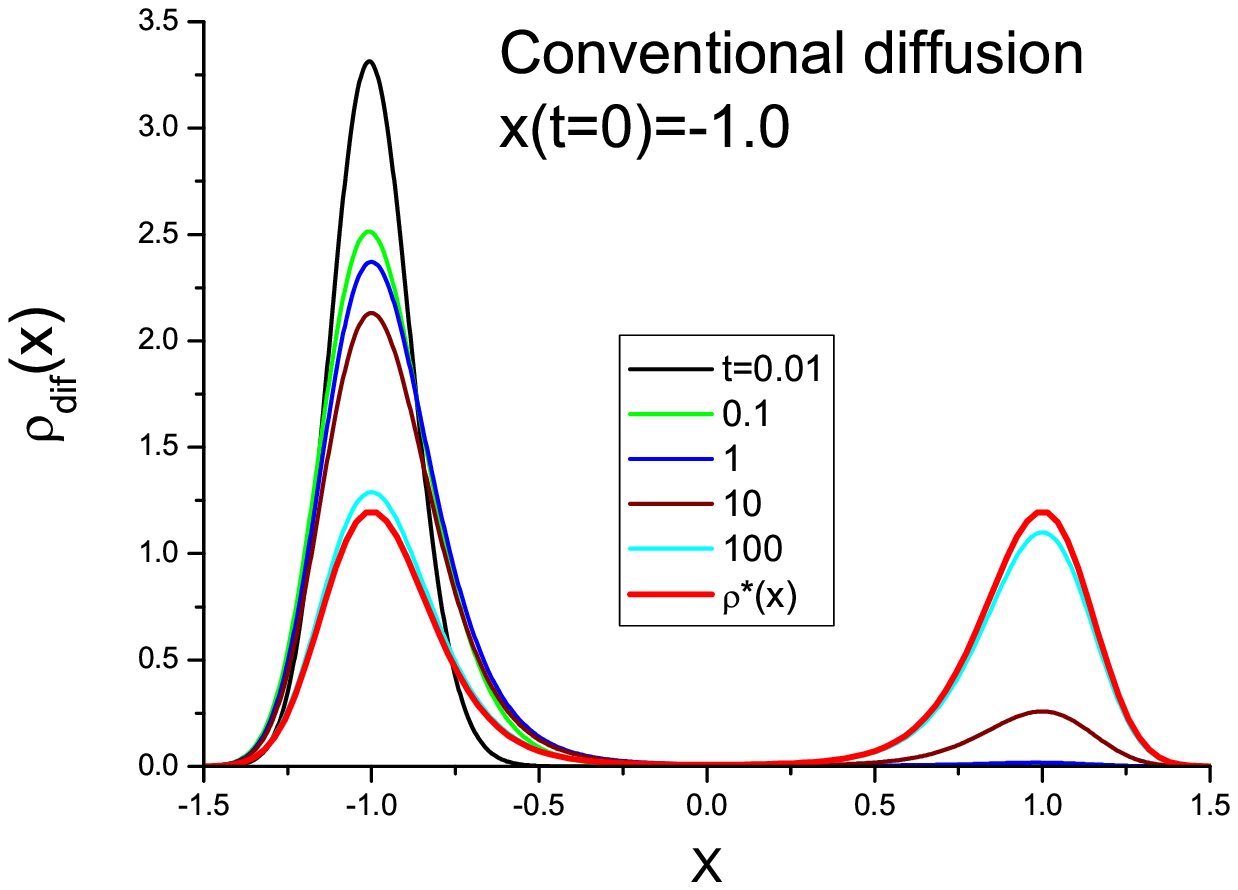}}
\hspace*{-0.5cm}
    \subfigure[]{\includegraphics* [width = 61mm]{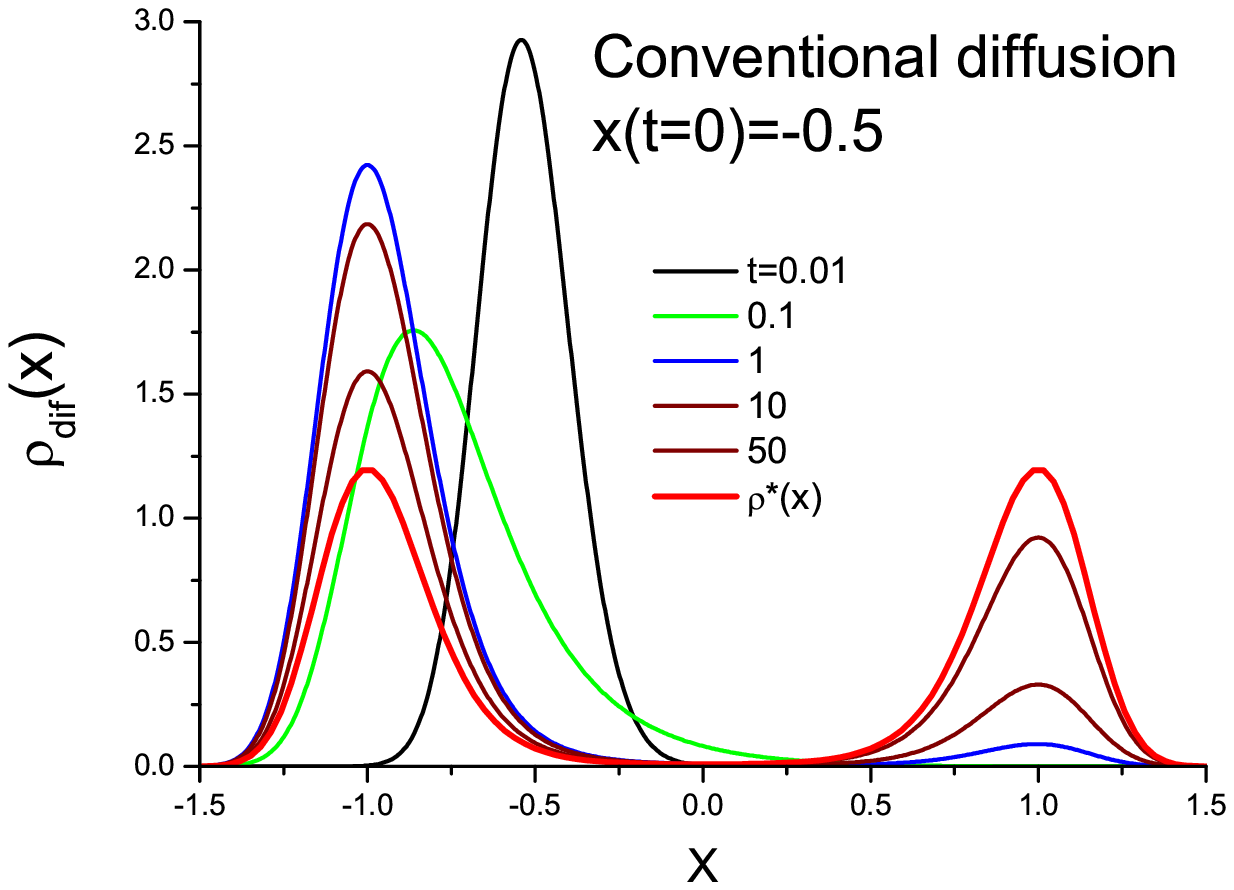}}
\hspace*{-0.5cm}
  \subfigure[]{\includegraphics* [width = 61mm]{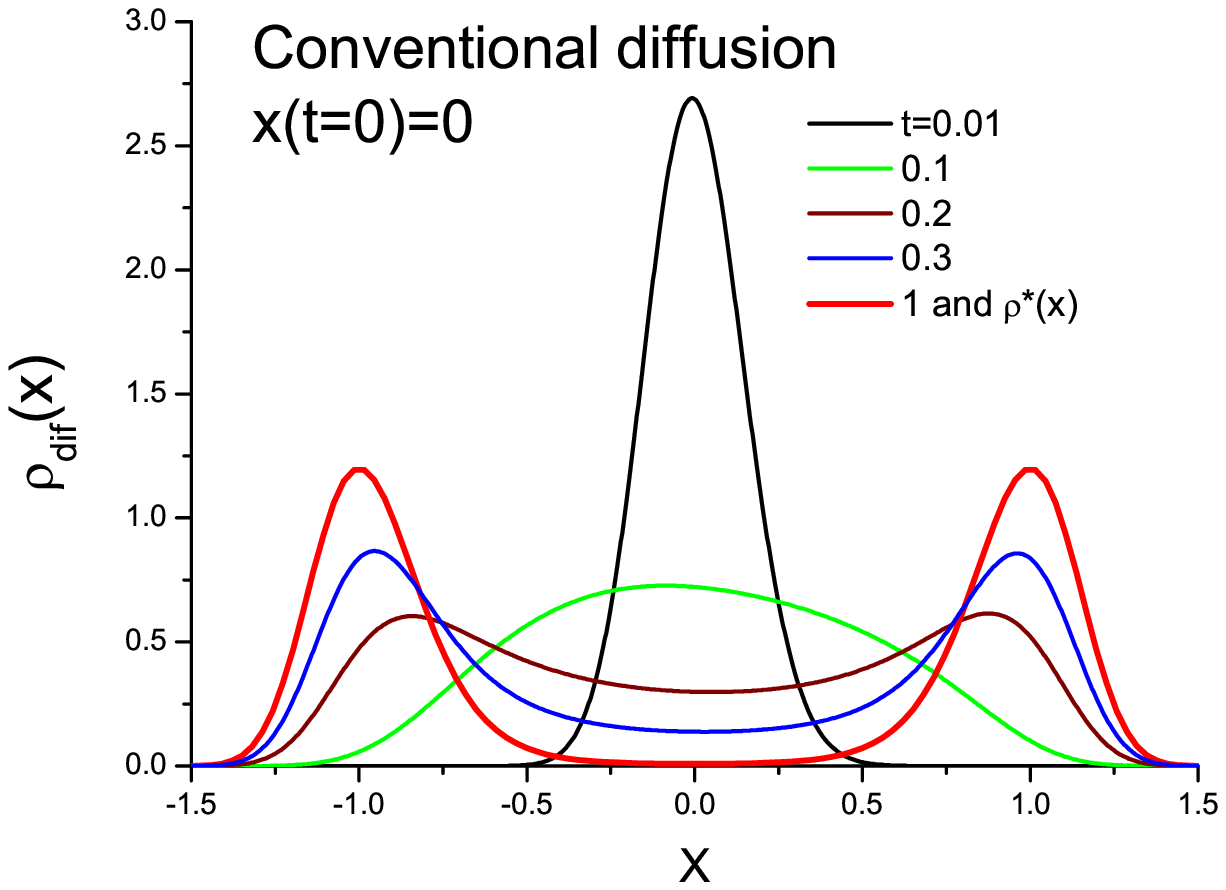}}
 \caption{The relaxation to the bimodal pdf  (\ref{zuz}) for three different type of processes: Langevin driven (panels (a) - (c)), semigroup driven (panels (d) - (f)) and
 Wiener driven (conventional FPE diffusion) (panels (g) - (i)). The initial Dirac delta-like pdf is located, respectively,   at $x=-1$ (panels (a), (d), (g)), $x=-0,5$ (panels (b), (e), (h)) and
$x=0$ (panels (c), (f), (i)). All three processes have different drift functions, corresponding to terminal pdf (\ref{zuz}) as indicated in the text. On panels (a) and (b), the log scale is utilized for better visualization of time evolution of initial unimodal pdf to the terminal bimodal one.} \label{fig:se}
\end{figure*}

\begin{figure*}
\begin{center}
\includegraphics [width=0.65\columnwidth]{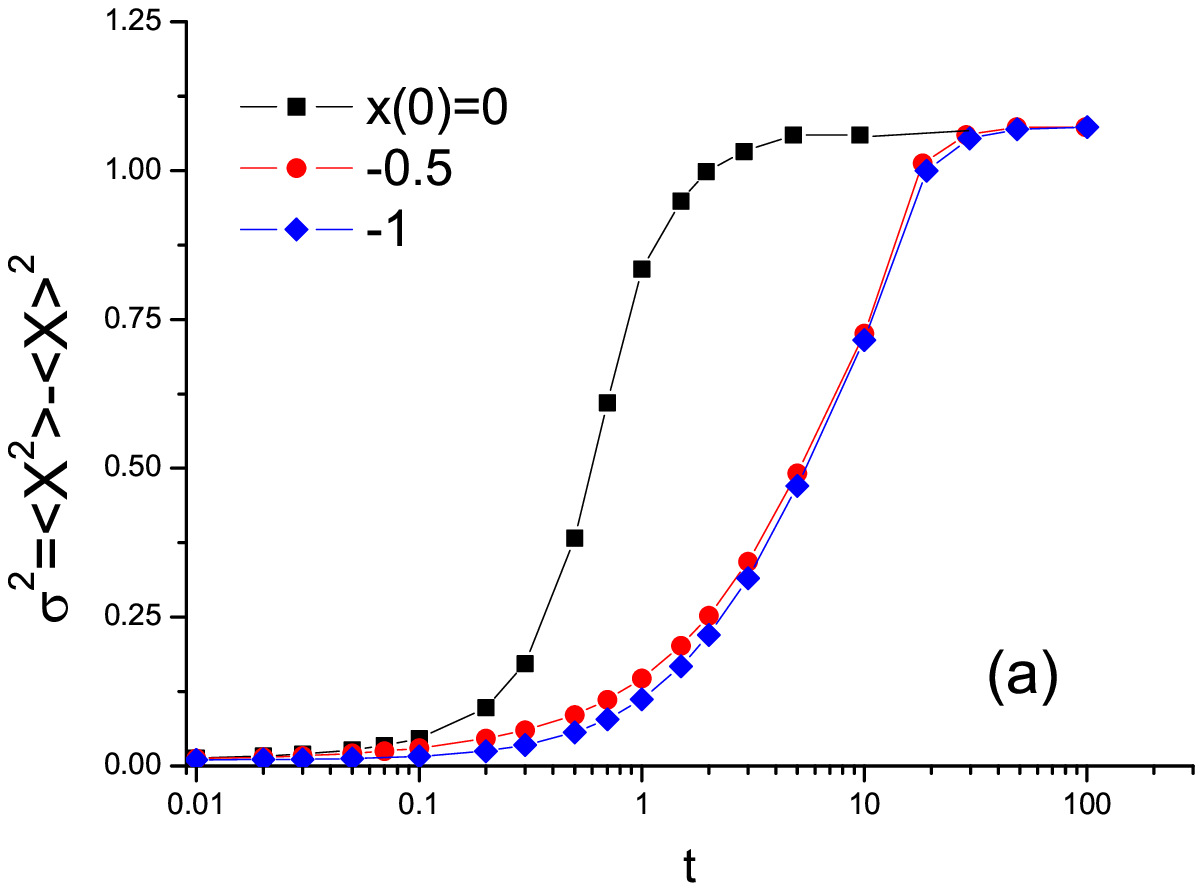}
\includegraphics [width=0.65\columnwidth]{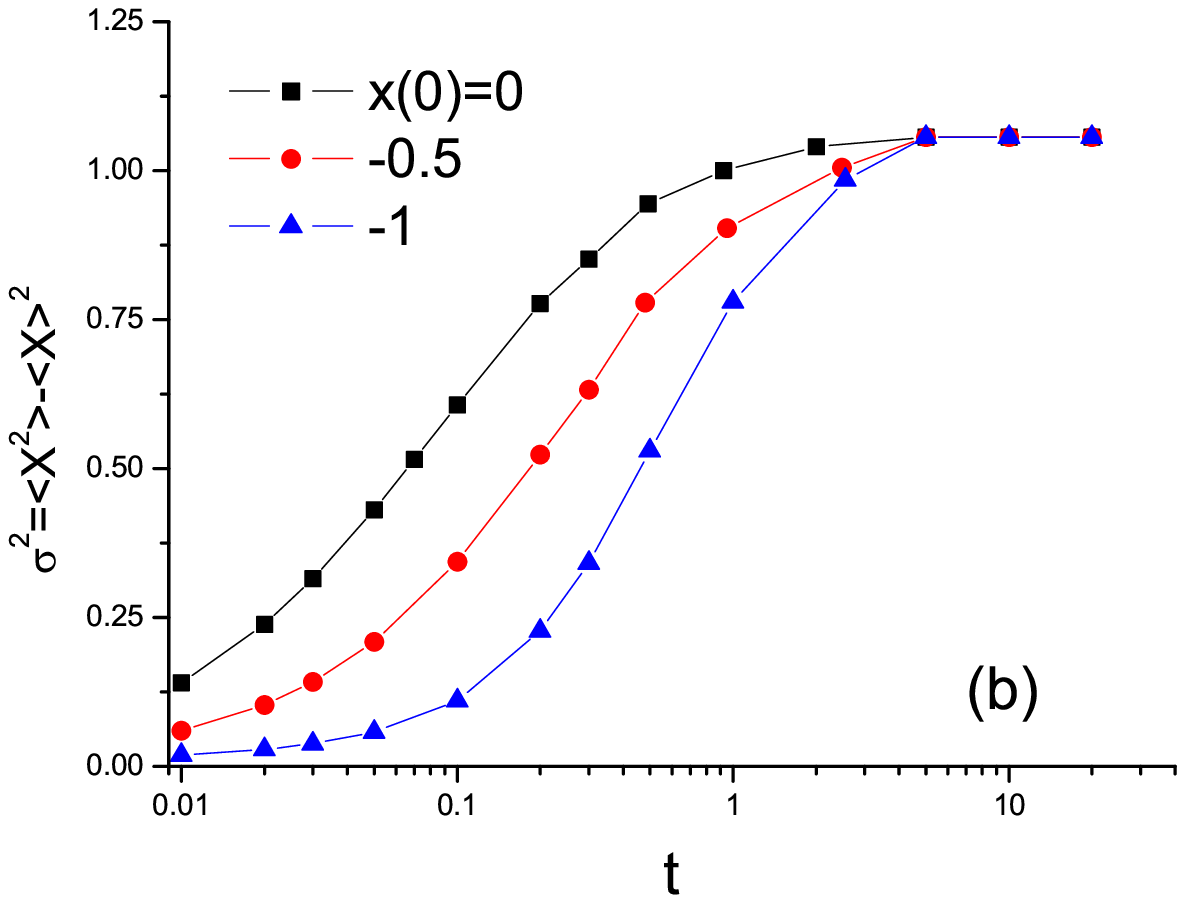}
\includegraphics [width=0.65\columnwidth]{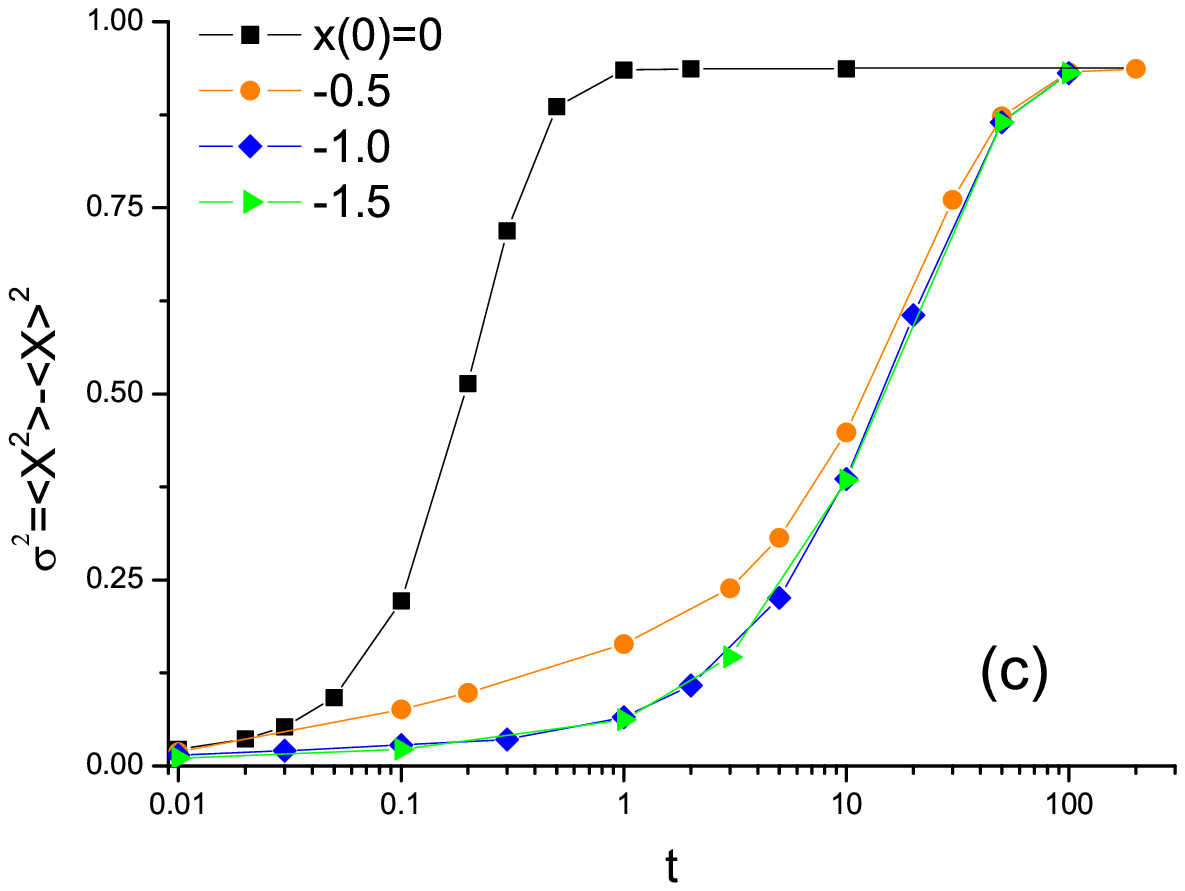}
\end{center}
\caption{The time evolution of variances  for pdfs in Fig. \ref {fig:se}. Panels (a), (b) and (c) correspond to Langevin driven, semigroup driven and conventional diffusion (Wiener driven) process
respectively}  \label{fig:os}
\end{figure*}

\subsubsection{Cauchy family  targeting}

 In relation to  the discussion of Section II,  our simulations have shown   that
  a  proper choice  of the  drift function in  the FPE, as $t \to \infty$  takes  the initial $\delta$-function
into   any     L\'{e}vy - stable distribution target,  or the like (e.g.  members of  the Cauchy family).
Accordingly,   conventional diffusion-type   processes  may in principle give rise to "heavy-tailed" asymptotic   pdfs.
Thus, heavy tails are not an exclusive property of L\'{e}vy processes.

 The only (and very substantial) difference between "normal" and fractional FPE  asymptotic outcome   is that in  the
 fractional case L\'{e}vy  flights may not relax to Gibbs-Boltzmann pdfs, \cite{klafter}. In turn,
 diffusion modeling  with  a suitable
 external  forcing admits a Gibbs form of asymptotic "heavy-tailed" pdfs.

Fig.~\ref{fig:x4}c reports an analogous (F-P dynamics vs MC simulation)   comparative procedure,  in case  when both Cauchy and Wiener drivers
are responding to  different  external forces, while giving  rise to a common asymptotic pdf.
Cauchy dynamics does not produce the Gibbs pdf.   However, the same pdf has the Gibbs form in the diffusive relaxation process.

We  have  directly  checked  that   a   target  function belonging to Cauchy family \eqref{ca1},
 $\rho_{*2}(x)$, is an asymptotic  outcome of the diffusion  process  via  path-wise simulation, with $b_{2,diff}(x)$ of Eq. \eqref{ca9}.
 We have confirmed that  the MC procedure  (with check-out  times $t=0,01$, $t=0,1$, $t=1$ and $t=8$)  correctly  reproduces
  the  qualitative picture  of the  time evolution  depicted in  Fig. \ref{fig:ca1}a. The ultimate outcome in the vicinity of the asymptotic
 pdf is reported in  Fig. \ref{fig:x4}c.

The very same reasoning may be adopted for  exemplary   jump-type processes, with Cauchy driver  and polynomial drift functions.
We note  that relaxation properties and confinement  of L\'{e}vy flights in various   external potentials  has
received attention in the literature, see  e.g. references in \cite{gs,gar,klafter,fogedby,chechkin,dubkov}.
For example, the Cauchy-Langevin dynamics has been analyzed for $\partial _t\rho = - |\nabla |\rho - \nabla (b\, \rho )$, with
$b_{jump}(x) =- x^3$. The corresponding  non-Gibbsian  invariant pdf reads $\rho _*(x)= 1/\pi (x^4-x^2+1)$, \cite{gs,dubkov}.

At this point we  can  again invoke   $\rho _{*2}(x)$, this time regarded as an asymptotic  target of the Cauchy-Langevin jump-type process with
the drift function
 $b_{2,jump}(x)= - x(x^2-3)/8$  of  section II.B.
Both the resultant    pdf  evolution  and the corresponding  Monte Carlo  simulation, in terms of  jump  sample  paths,
 give consistent outcomes.

Concerning  mathematical subtleties of the Cauchy semigroup dynamics, their well-definiteness comes from the properties of
semigroup potentials. This issue has received due attention in Ref.~\cite{olk},
see also \cite{gs,gar}.

\subsection{Relaxation time rates}

In recent years there has been continually growing interest in various theoretical  random walk models, clearly
motivated by the inefficiency of the standard Gaussian modeling paradigm. That refers to various sub-fields of physics, extending to
 chemistry, biology, biophysics  and  financial mathematics.

 The classical concept of Brownian motion   pervades the  whole theory of stochastic
processes.  The pdf of a homogeneous Brownian motion solves
 a   Fokker-Planck  (actually, heat) equation and  remains sweeping unless confined in a finite region or subject to external forces.
Typically  the time-evolving  pdf, that is
initially concentrated  at  (or about) a point, with the flow of time takes the Gaussian form, whose width  grows in time as $t^{1/2}$.
 This  diffusion processes  is called the normal diffusion.

In  a broad  field of anomalous diffusions  (and more general non-Gaussian processes), whose microscopic modeling  may involve
both Markovian and non-Markovian random dynamics,   another   property has been discovered:    $<X^2(t)>  \sim  t^{\gamma }$,  with $0<\gamma <2$.
 This dynamics is sweeping as well, hence precluding the existence of any well
defined relaxation pattern (except for a possible relaxation to a uniform distribution, if the motion is spatially confined).

For a special class of non-Gaussian jump-type processes,
first and second moments of the pdf may not exist at all.
 That happens in case   L\'{e}vy - stable distributions.
Due to  long tails of the pdf, in the least   their  second (and all  higher) moments are nonexistent.
The existence of the first moment is granted  only for a suitable subclass.

In the presence of external potentials  the confinement (taming) of L\'{e}vy flights may occur,  and   quite in
affinity with standard diffusion  processes in external force fields,  asymptotic invariant pdfs  may be approached.  The speed (time rate)
 of the corresponding relaxation  processes is worth addressing  and set against that arising in the sweeping motion.

Somewhat intriguing point in the anomalous transport  is whether,  often used terms
like  "subdiffusion" or "superdiffusion" may at all  have  meaning in   connection with  L\'{e}vy flights,
 under confining  (i.e. relaxation)  regimes.
Clearly,  there always appears a transitional period  during which the diffusion,  jump-type or semigroup-driven
 process  may be analyzed in terms
 of a time dependent variance (unless non-existent). This property is regarded as  generic in  relaxing to equilibrium
  diffusion processes.

Our purpose is to check whether one may expect any regularity in the time evolution of variances for diffusion-type and jump-type (Langevin and semigroup-driven) processes. That could  possibly  give  useful hints towards their hierarchical  classification  with respect to relaxation time rates.  May we talk about "slow", "fast" or "super-fast" processes at all ?

Such hypothesis might look plausible on the basis of our previous analysis of  relaxation scenarios whose asymptotic targets were pdfs from the Cauchy family, Eq.~\eqref{cf}.
However, the situation is not that simple and appears to be incongruent with naive expectations. This issue we shall discuss below.

\subsubsection{Relaxation to the Cauchy pdf $\rho _{*1}$}
Our first  observation pertains to the Ornstein-Uhlenbeck-Cauchy process (OUC, Langevin-driven)  and its semigroup-driven relative with a common target pdf $\rho _{*1}$,  \cite{gs1}. 
It is seen form Fig. \ref{fig:ca1}, that both the transitional behavior and the time needed to reach  (almost,  within the figure resolution limits) the asymptotic pdf, indicate that the semigroup dynamics is running  somewhat  slower as  compared to Cauchy-Langevin-driven  dynamics.
Indeed, the transitional period of motion admits a direct analysis of the  half-widths  (second moments of $\rho _{*1}$ are non existent, see above) time rate in terms of
respective exponents of  $ t^{\gamma }$. Namely, for the  proper OUC process, the transitional exponent equals
 $\gamma = 0.58$, while in the semigroup-driven  case, we have $\gamma = 0.45$.

Both processes are of the jump-type and have no second moments. We note that the Cauchy-Langevin driving sets the  $\rho (x,t)$  in the vicinity of the invariant one after time $t=8$, while the semigroup-driving needs $t=15$ to the same end.  The diffusion scenario, according to Fig.~\ref{fig:ca1}c sets at $\rho _{*1}$ after time $t=5$.

\subsubsection{Relaxation to the bimodal pdf}

Now we  discuss  various aspects of relaxation time rates for diffusion-type, Cauchy-Langevin driven and  Cauchy semigroup-driven processes. All of them are started from  "almost" Dirac delta pdfs (albeit  localized about  three  different initial points, and all terminated
 at an asymptotic  bimodal  pdf, \cite{gs1}
\begin{equation}\label{zuz}
  \rho _*(x) = {\frac{2a(a^2 +b^2)}{\pi }} {\frac{1}{(a^2+b^2)^2 + 2(a^2 -b^2)x^2 + x^4}}
  \end{equation}
Here, parameters   $a$ and $b$ are, respectively, real and imaginary parts of complex roots of the  cubic equation  $z^3 +z - 1/4=0$.
 Their approximate values are  $a\simeq 0.118366$ and $b\simeq 1.0208$.

This  exemplary bimodal pdf  is an asymptotic distribution of the Cauchy-Langevin driven dynamics
 of $\rho (x,t)$ with the drift
$b_{jump} = - \nabla  V_*(x)$,  where  $V_*(x)= (x^2 -1)^2$.
The Cauchy-Langevin    evolution of $\rho (x,t)$, that is started from three
 different locations $x=0$, $x=-0,5$ and  $x=-1$
 in  the double well potential  $V_*$,  is depicted in Fig.~ \ref{fig:se}, panels (a)-(c).

The data collected for Fig.\ref{fig:se} (a)-(c) allow to deduce the time-dependence of variances, in the
transitional regime.  They display  $\sim t^{\gamma }$ behavior which  depends on the initial  pdf location data.
Namely, for  $x=-1$ and  $x=-0,5$ we have respective exponents  $\gamma = 0.53$, while $x=0$
yields $\gamma =1$.

The Cauchy  semigroup-induced evolution of $\rho (x,t)$, has been simulated  under the very same (as in the
 Cauchy-Langevin case, Fig.\ref{fig:se} (a)-(c)) initial data. The corresponding patterns of behavior are depicted in Fig.\ref{fig:se}, panels (d)-(f).
In Fig.~\ref{fig:os} we report the time evolution of respective variances. The $\sim t^{\gamma }$ behavior
in the transitional regime is characterized by exponents: $\gamma =1$ for $x=-1$, $\gamma = 0.53$
for $x=-0.5$  and $\gamma =0.2$ for $x=0$.

 For completeness we have performed analogous simulations with the Wiener driver in action, i.e. for the conventional  diffusion-type  process that interpolates between the common (for all three types of processes) initial data and the terminal bimodal pdf, see e.g. also Ref. \cite{sokolov}.
  The results are shown on panels (g) - (i) of Fig.\ref{fig:se}.
In Fig.~\ref{fig:os} we  also  report the temporal behavior of  related  variances. The $\gamma $ exponents in the
transitional regime read: for $x=-1.5$ and $x=-1$ we get $\gamma =0.75$,  $x=-0.5$ corresponds to
$\gamma =0.5$, while $x=0$ to $\gamma =1.2$.

The preceding analysis shows that standard classification of  anomalous  diffusion processes as "subdiffusions"  or   "superdiffusions" on the basis of
the exponent $\gamma$,  is invalid in the presence of  external forces.  Moreover, this  exponent strongly depends on both the
particular   localization  of the initial pdf  (effectively, an initial  position of a fictitious particle),
 with respect to the corresponding potential profile,  and on the potential  curvature.

\section{Conclusions}

The main message of the present paper is that, under suitable
confinement conditions,
 the ordinary Fokker-Planck equation may  generate a non-Gaussian
 heavy-tailed pdf (like e.g.  Cauchy or  more  general  L\'{e}vy stable distribution)  as its
long time asymptotics.  That implies a continuous interpolation
between  an  initial highly localized pdf (like Dirac $\delta$ -
function) having all  moments and the terminal  heavy - tailed pdf,
 with only few (or none at all) moments in existence.

  Since it is  the fractional generalization of FPE  which is customarily invoked to generate the same heavy-tailed  pdfs,  albeit
   with  a  very  different  choice of drift functions, we
 have uncovered   an unexpected link between non-Gaussian  jump-type processes (inherent in  the fractional FPE derivation)
 and ordinary  diffusion processes that are based on the Gaussian (Wiener driver) paradigm.

In the present paper we have paid attention not only to random  processes that have found their place in the literature,
all stemming from different forms of  Langevin modeling (here, only additive noise has been considered), but also to alternative
jump - type scenarios based  on the concept of the semigroup-driven dynamics, \cite{gs,olk}. This pattern of dynamical behavior has not been
satisfactorily explored within the area of anomalous random transport.

To  make our  findings  transparent, we have undertaken a   direct
numerical  verification  of  them,  by invoking two kinds of
modeling. The first one amounts to  solving  numerically  both
ordinary (with the  Laplace operator) and fractional FPE's  with the
drift functions  selected to  generate, the   same for both,
non-Gaussian (specifically Cauchy family \eqref{ca1}) pdfs. The
second one presumes that one has the  above drift functions at
disposal. Then, we undertake a direct numerical modeling of  random
trajectories  (sample paths)  of the underlying stochastic process,
both Brownian and of the  jump-type.  These trajectories have  next
been used to generate the associated pdfs which   turn out to
coincide with  those obtained from numerical solution of both
ordinary and fractional FPE's.

We note that a traditional  interpretation of random data relies on the existence of at least two lower moments of involved probability distributions.
Therefore, the case of L\'{e}vy flights could have been  placed  within this interpretational  paradigm  only under confining conditions.
In such case, in view of  the   existence of second moments, one may  address their time-dependence in transitional regimes, when the process is yet far from
its equilibrium pdf. The above numerical solutions of the FPE and  of  its fractional analog permit us to investigate the time dependence of different moments (see Fig. 8) of the corresponding pdfs.

At this point we  have passed  to an  approximation of the variance $\sigma ^2$ by the power law $\sigma^2 \sim t^\gamma$,
 with the hope that its validity  would  shed some light on the problem
of assigning terms like "subdiffusion" or "superdiffusion" to anomalous transport processes.   Namely, the time evolution of variances is often
considered as a signature of subdiffusive, normal  or superdiffusive behavior, typically considered with no bounds on the duration time of the
process (unless set be experiment).  Taken literally, all such motions would belong to the sweeping category, where the variance grows indefinitely.

In this connection, our  analysis shows that (i)  near initial
$t \to 0$  the exponent $\gamma$ depends strongly on time,  (ii) deeply  in the  asymptotic  $t \to \infty$ regime $\gamma$  can be equated to $0$,
  (iii)  in a transitional
 regime (i.e. for pdfs lying somewhere   between initial and terminal ones) the exponent $\gamma$    happens to take
 almost constant \it positive \rm   values  (observe e.g.   almost straight lines in time dependence of variances in
 the middle parts of  the corresponding plots
 in Fig. 8).

We have focused on possible signatures of (ab)normal relaxation patterns,  expecting that  in  a transitional regime we might possibly
identify an undoubtful subdiffusive or  superdiffusive  dynamics.
 Our  answer is negative. We have confirmed  that widely used terms  "sub-" and/or "superdiffusion"   are rather vague and
  become meaningless for confined stochastic processes,
 likewise in the presence of  L\'{e}vy and Wiener drivers. The   pertinent  exponent value is not indicative for
 the stochastic process in question, to justify its naming (sub-, normal, super-), because it  depends strongly on the
  location (position)  of the initial  pdf relative to the drift  potential profile,
  and on its curvature.

It appears that the  above  "anomaly" in the description of random motions appears  to have much  deeper foundations.
 One  can alternatively use  incompatible microscopic mechanisms   to model an engineered asymptotic approach to  the same a
  priori prescribed non-Gaussian    target  pdf.

 If those pdfs are   an outcome of a statistical analysis of   experimental data, clearly a  distinction between diffusive
  and jumping patterns of  dynamical behavior    appears not to be   that sharp as commonly expected.
  On the other hand, a  possible discrimination tool, e.g. the single-particle/molecule
   experiments in the nano- or mesoscopic  domain  should be set under scrutiny,    since   microscopic features of motion
may not be verifiable at current levels of precision/resolution in collecting the data.
   The specific  stochastic model that would   seemingly  fit to the data, might possibly  be not
 more than a lucky guess, to be invalidated in a more thorough analysis.

Another message worth spelling   is  that  apparently divorced dynamical
mechanisms  (Langevin-Cauchy, Cauchy semigroup and diffusion-type process) may share common  asymptotic pdfs.
 Contrary to the  wide-spread belief,   heavy-tailed  probability distributions,
casually  associated with jump-type processes, may as well have  Gaussian  (e.g. diffusive)  origins.  That is possible
 if drift functions refer to   properly functioning potentials, which we attribute to a cumulative effect of  inhomogeneities
   of the environment). Their  role would be  to    \it attenuate \rm the  casually considered    counterbalancing  of the  Wiener driver, "normally"  necessary to  eliminate   heavy tails of the   resultant pdf.

\end{document}